\documentclass[twocolumn]{aastex63} 

\usepackage{graphicx}
\accepted{June 17, 2021}

\begin{document}

\title{Speckle Observations of TESS Exoplanet Host Stars. II. Stellar Companions at 1-1000~AU \\ and Implications for Small Planet Detection}

\author[0000-0002-9903-9911]{Kathryn~V.~Lester}
\affil{NASA Ames Research Center, Moffett Field, CA 94035, USA}

\author[0000-0001-7233-7508]{Rachel~A.~Matson}
\affiliation{U.S. Naval Observatory, Washington, D.C. 20392, USA}

\author[0000-0002-2532-2853]{Steve~B.~Howell}
\affil{NASA Ames Research Center, Moffett Field, CA 94035, USA}

\author[0000-0001-9800-6248]{Elise~Furlan}
\affiliation{NASA Exoplanet Science Institute, Caltech/IPAC, Pasadena, CA 91125, USA}

\author[0000-0003-2519-6161]{Crystal~L.~Gnilka}
\affil{NASA Ames Research Center, Moffett Field, CA 94035, USA}

\author[0000-0003-1038-9702]{Nicholas~J.~Scott}
\affil{NASA Ames Research Center, Moffett Field, CA 94035, USA}

\author[0000-0002-5741-3047]{David~R.~Ciardi}
\affiliation{NASA Exoplanet Science Institute, Caltech/IPAC, Pasadena, CA 91125, USA}
	
\author[0000-0002-0885-7215]{Mark~E.~Everett}
\affiliation{NSF’s National Optical-Infrared Astronomy Research Laboratory, Tucson, AZ 85719, USA}

\author[0000-0003-4236-6927]{Zachary~D.~Hartman}
\affiliation{Lowell Observatory, Flagstaff, AZ 86001, USA}
\affiliation{Department of Physics \& Astronomy, Georgia State University,  Atlanta GA 30303, USA}

\author[0000-0001-8058-7443]{Lea~A.~Hirsch}
\affiliation{Kavli Institute for Particle Astrophysics and Cosmology, Stanford University, Stanford, CA 94305, USA}

\correspondingauthor{Kathryn Lester}
\email{kathryn.v.lester@nasa.gov}

\begin{abstract}

We present high angular resolution imaging observations of 517 host stars of TESS exoplanet candidates using the `Alopeke and Zorro speckle cameras at Gemini North and South. The sample consists mainly of bright F, G, K stars at distances of less than 500~pc. Our speckle observations span angular resolutions of $\sim$20~mas out to 1.2~arcsec, yielding spatial resolutions of $<$10 to 500~AU for most stars, and our contrast limits can detect companion stars $5-9$ magnitudes fainter than the primary at optical wavelengths. We detect 102 close stellar companions and determine the separation, magnitude difference, mass ratio, and estimated orbital period for each system. Our observations of exoplanet host star binaries reveal that they have wider separations than field binaries, with a mean orbital semi-major axis near 100~AU. Other imaging studies have suggested this dearth of very closely separated binaries in systems which host exoplanets, but incompleteness at small separations makes it difficult to disentangle unobserved companions from a true lack of companions. With our improved angular resolution and sensitivity, we confirm that this lack of close exoplanet host binaries is indeed real. We also search for a correlation between planetary orbital radii vs. binary star separation, but given the very short orbital periods of the TESS planets, we do not find any clear trend. We do note that in exoplanet systems containing binary host stars, there is an observational bias against detecting Earth-size planet transits due to transit depth dilution caused by the companion star.

\end{abstract}


\section{Introduction} 
The Kepler, K2, and TESS missions have provided precise, light curve photometry for millions of stars and contributed breakthrough advancements in many fields of stellar astrophysics \citep{borucki10, howell14, ricker15}. This includes the discovery of thousands of extra-solar planets through the transit method, where a planet passing in front of a star causes a periodic dip in brightness. Due to the large pixel sizes of 4" for Kepler/K2 and 21" for TESS, nearby stars or unresolved stellar companions within the aperture dilute the transit depth and bias the measured planet radius \citep{ciardi15}. About 50\% of solar-type exoplanet hosts have companions \citep[e.g.,][]{horch14, matson18}, so follow-up high resolution imaging is needed to search for unresolved companions and aid exoplanet validation and characterization.

Understanding how binary companions affect the formation, evolution, and survival of exoplanets is also an important component to understanding planet formation overall. Theoretical studies show that a companion can truncate the protoplanetary disk, leaving less material for planets to form \citep{martin14, jc15}, cause the migration of gas giant planets \citep{dawson18}, or disperse the disk before planets are able to form \citep{cieza09, kraus12}.  Observational studies using high resolution imaging \citep{kraus16, fontanive19, ziegler20, ziegler21, howell21} and radial velocities \citep{wang14, hirsch20} have found fewer companions within 100~AU around exoplanet hosts than around field stars, supporting the idea that close stellar companions do suppress planet formation. 

However, it is difficult to disentangle unobserved companions from a true lack of companions without sufficient angular resolution. The angular separations of the closest companions ($<50$~AU) around Kepler exoplanet hosts are near or below the diffraction limit of most telescopes, but TESS observed brighter, nearby stars for which we can resolve companions at smaller physical separations from the host star. We present the highest angular resolution speckle images of TESS exoplanet host stars in search of companions at $1-100$~AU where planet formation would be greatly hindered. We describe our observations and companion detections in Section~\ref{sect-obs}, then create a simulated sample of binary stars to investigate the separation distribution of exoplanet host binaries in Section~\ref{sect-sim}. Our results and conclusions are presented in Sections \ref{sect-results} and \ref{sect-con}, respectively.

\section{Observations}\label{sect-obs}
\subsection{Speckle Imaging}
\newcommand{\numobstoi}{517 }    
\newcommand{\numobsexo}{412 }  
\newcommand{\numobsfp}{105 }     
\newcommand{\numobsFGK}{315 }

We observed \numobstoi stars from the TESS Objects of Interest (TOI) catalog using the `Alopeke and Zorro speckle cameras \citep{scott18, scott21} on the Gemini 8.1~m North and South telescopes respectively from May 2019 to December 2020. Before each observing season, targets were selected from the latest version of the TOI catalog of likely planet candidates. All of our observations are listed in Tables~\ref{northlog} and \ref{southlog} in the Appendix, with the TOI and TIC numbers, TESS magnitude, UT date, the inverse of the parallax from Gaia EDR3 \citep{gaia, gaiaEDR3} for the distance, effective temperature of the host star from the TESS Input Catalog \citep[TIC v8.1,][]{tic81}, and notes for each target. The Julian date of each observation can be found in the headers of the archival data hosted on the Gemini Observatory Archive\footnote{\href{https://archive.gemini.edu/searchform}{https://archive.gemini.edu/searchform}}. At least three image sets were obtained for each target, where one set consists of 1000 60~ms exposures taken in a 562~nm and an 832~nm filter simultaneously. Additional image sets were taken for fainter targets ($V>9$ mag), and a point source standard star was observed immediately before or after each target for calibration. 

\begin{figure*}
\centering
\includegraphics[width=0.85\textwidth]{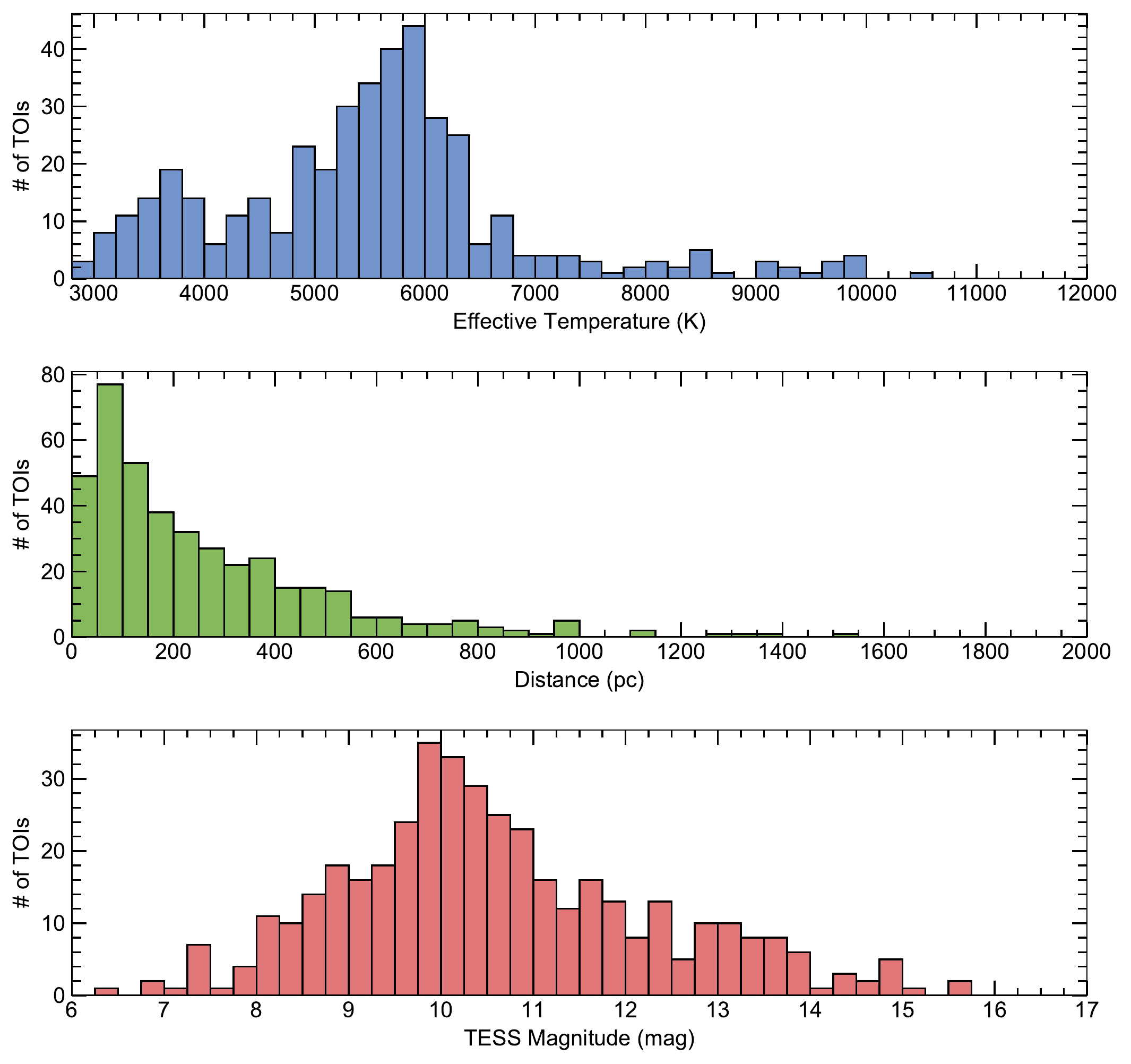}
\caption{Stellar parameter distributions for the \numobsexo exoplanet host TOI's in our sample, including stellar effective temperature (top), Gaia EDR3 distance (middle), and TESS magnitude (bottom).  \label{samplehist}}
\end{figure*}

The data were reduced using the pipeline developed by the speckle team \citep{howell11, horch11} to calculate and average the power spectrum of each image, then correct for the speckle transfer function by dividing the mean power spectrum of the target by that of the standard star. The pipeline also produces a reconstructed image of each target with a field of view of 1.3" in radius around the target. The contrast limits for each target were determined with the method described in \citet{horch11}, which uses the background flux levels to determine the faintest companions one could reliably detect at each separation. These $5\sigma$ contrast limits at 0.2\arcsec\ and 1.0\arcsec\ in the blue and red filters are also listed in Tables~\ref{northlog} and \ref{southlog}.

Due to the large TESS pixels, some transit detections were identified to be false positives through follow-up photometry. If the light curve shows chromaticity or v-shaped dips of alternating depth, the event must be a binary eclipse rather than a planet transit, then confirmed with follow-up observations to identify the eclipsing source within the TESS aperture. The transits for \numobsfp of the TOI's in our sample have been identified as false positives (or false alarms) on the Exoplanet Follow-up Observing Program (ExoFOP)\footnote{\href{https://exofop.ipac.caltech.edu/tess}{https://exofop.ipac.caltech.edu/tess}} website, noted in Tables~\ref{northlog} and \ref{southlog}. We have eliminated these host stars from further consideration here, leaving us with \numobsexo stars. A few additional systems were classified as ambiguous planetary candidates (APC), for example, if only a single transit was observed. Because these systems have not yet been confirmed as false positives, we kept them in our sample and discuss their impact in Section~\ref{results2}.

Figure~\ref{samplehist} shows the distributions in effective temperature, distance, and TESS magnitude for our \numobsexo exoplanet host TOI's. Our sample contains primarily bright, Solar-type stars with smaller distances from Earth compared to Kepler host stars, allowing Gemini speckle observations to detect companions quite close to the target star. For distances of $100-500$~pc, our inner angular resolution of 0.017" at 562~nm corresponds to separations of $1.7-8.5$~AU, while our inner angular resolution of 0.026" at 832~nm corresponds to separations of $2.6-13.0$~AU. Therefore, we can successfully detect stellar companions in the $1-100$~AU regime and determine their properties.

\begin{figure}
\centering
\includegraphics[width=0.45\textwidth]{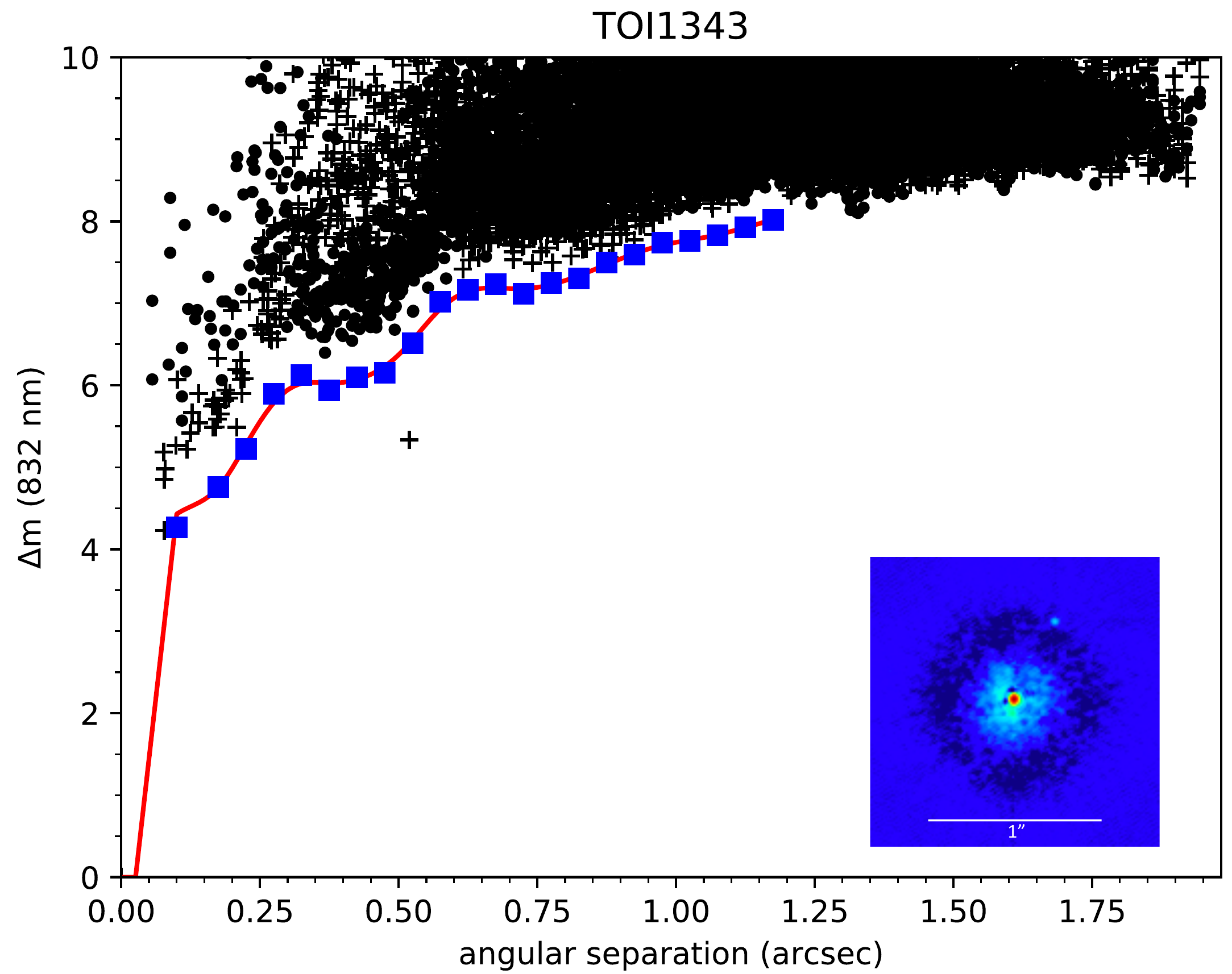}
\caption{Example detection limit as a function of radius from the center for a binary star (TOI 1343). The black points represent the local maxima (crosses) and minima (dots). The blue squares mark the $5\sigma$ background sensitivity limit within $0.05"$ bins, and the red line corresponds to a spline fit. A companion is detected if any maxima points lie below the detection limit, such as the point near 0.5" in this example. The inset plot shows the reconstructed image where the companion is visible in the top right (northwest) corner. 
\label{example}}
\end{figure}

\break

\subsection{Detected Companions}
\newcommand{\numcomp}{102 }       
\newcommand{\numsys}{99 }            
\newcommand{\numcompexoall}{73 } 
\newcommand{\numcompexo}{58 }      
\newcommand{\numcompexofgk}{47 } 
\newcommand{\numcompfp}{29 }    

Binary stars produce a characteristic interference fringe pattern in the power spectrum from which the companion's angular separation ($\rho$), position angle ($\theta$, measured East of North), and magnitude difference ($\Delta m$) can be determined from a weighted least squares fit. Figure~\ref{example} shows an example contrast curve and reconstructed image for the binary star TOI 1343. We detected a total of \numcomp stellar companions around \numsys TOI's in our sample: \numcompexoall companions around exoplanet host stars and \numcompfp companions around transit false positive stars. Our results are listed in Table~\ref{companiontable} for both the exoplanet candidate host stars and the false positive stars, with the TOI number, observation date, and companion parameters ($\rho, \theta, \Delta m$) determined from both the 562~nm and 832~nm images. We detected some companions only in the 832~nm filter, and could not constrain the magnitude difference for three of these systems. Results from multiple nights were averaged together because no orbital motion could be seen, so any TOI's with duplicate entries in Table~\ref{companiontable} were found to be triple systems. The typical uncertainties in the separation, position angle, and magnitude difference are 0.01", 0.5~degrees, and 0.5~mag, respectively. 

Several of these companions were previously detected by the NESSI speckle camera on the WIYN 3.5~m telescope \citep{howell21} or HRCam on the SOAR 4.1~m telescope \citep{ziegler20}. Most separations and position angles from Gemini agree with those from WIYN and SOAR at the $1\sigma$ level. One exception was TOI 1191, for which a companion was found at $\rho=0.68$" with $\Delta m=5$~mag by `Alopeke but at $\rho=0.046$" with $\Delta m=1.54$~mag by NESSI. Palomar adaptive optics observations also found a companion at 0.5", consistent with the `Alopeke results. This TOI could be a triple system, but future observations at higher signal-to-noise are needed to confirm these companions. 

Due to speckles becoming de-correlated at 1.2", companions beyond this limit have larger uncertainties, especially in $\Delta m$. These wide companions are also more likely to be line-of-sight companions, rather than gravitationally bound systems \citep{horch14, matson18}, but this can be confirmed with Gaia proper motion and parallaxes \citep[e.g.,][]{mugrauer20}. We do not consider these wide companions nor the companions around transit false positive TOI's in the rest of our analysis. Figure~\ref{companions} shows the magnitude difference versus angular separation for the remaining \numcompexo companions with $\rho < 1.2"$ as well as our typical detection limit. We also chose to include the companions detected at WIYN in our analysis to increase our sample size; \citet{howell21} observed an additional 130 Solar-type TOI's and detected 18 companions with separations less than 1.2" that are included in our comparison of exoplanet host binaries and field binaries in Section~\ref{results2}. This final sample only includes binaries, as no triple systems with separations less than 1.2" were detected.

\subsection{Binary Properties}
\begin{figure*}
\centering
\includegraphics[width=0.49\textwidth]{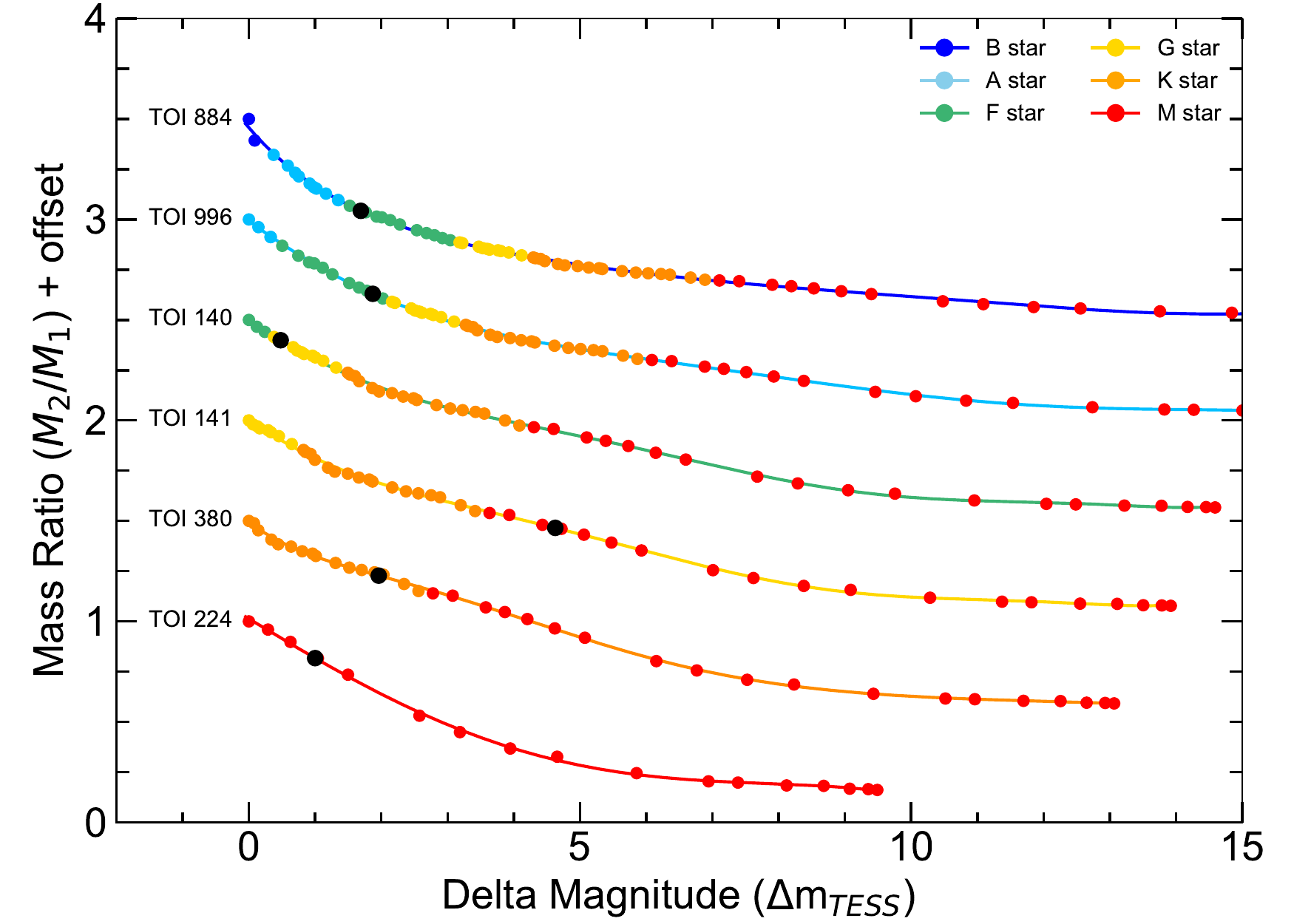}
\includegraphics[width=0.49\textwidth]{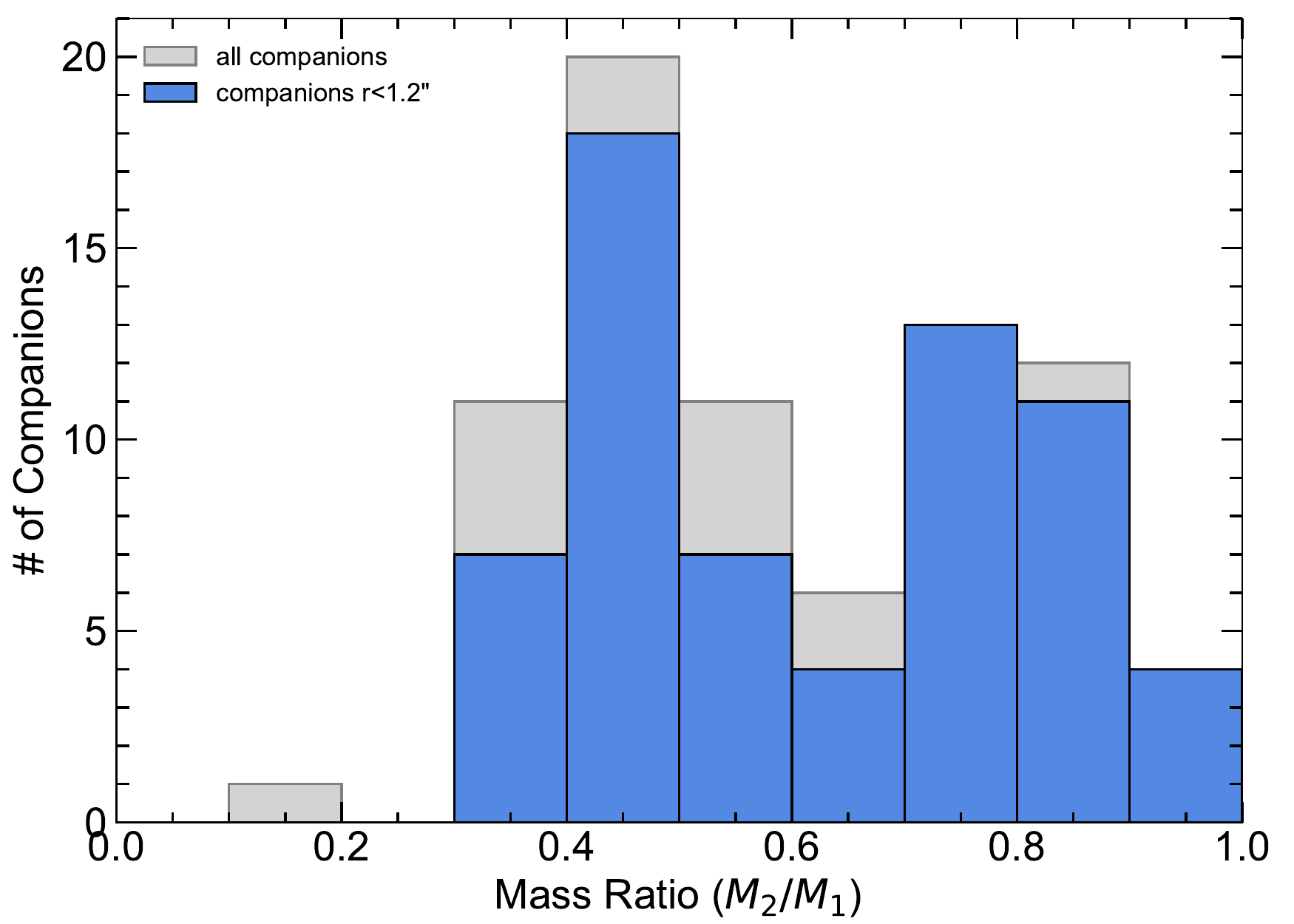}
\caption{\textit{Left:} Mass ratio as a function of magnitude difference for example exoplanet host binaries. The \citet{pecaut13} model values are shown as the colored dots and the polynomial fit for each binary is shown as the solid line; both are color-coded by spectral type of the possible companions and primary star, respectively. The values of the observed companions are shown as the solid black dots. \textit{Right:} Histogram of the mass ratio for exoplanet host stars. The light grey histograms represent all detected companions, while the blue histograms represent companions with separations $<1.2"$.
\label{massratio}}
\end{figure*}

We estimated the mass ratio ($M_2/M_1$) of each detected binary system using the method in \citet{matson19} and \citet{howell21}. For each primary star, we found the nearest value in the Modern Mean Dwarf Stellar Color and Effective Temperature Sequence \citep{pecaut13} based on the effective temperature, then calculated the $V$-band magnitude differences ($\Delta m_V$) of all possible companions (stars with cooler effective temperatures). For companions detected in our 562~nm filter, we fit a polynomial to the mass ratio as a function of $\Delta m_V$ then calculated the mass ratio corresponding to the observed magnitude difference. For companions detected in the 832~nm filter, we converted the magnitude differences from the $V$-band to the TESS band ($\Delta m_{TESS}$) using the relations in \citet{stassun18}, then interpolated the mass ratio from the polynomial fit. We took the average if a companion was detected in both filters, then estimated a typical uncertainty of 0.02 from the difference in the mass ratio across filters. Our results are listed in Table~\ref{companiontable}, and Figure~\ref{massratio} shows the mass ratio as a function of magnitude difference for select binaries and a histogram of the mass ratios. Our sample of exoplanet host binaries does not show the peak towards $M_2/M_1 = 1$ seen in field binaries, possibly because our binaries have long orbital periods that do not show a preference for equal mass components \citep{raghavan10}.

\begin{figure*}
\centering
\includegraphics[width=0.49\textwidth]{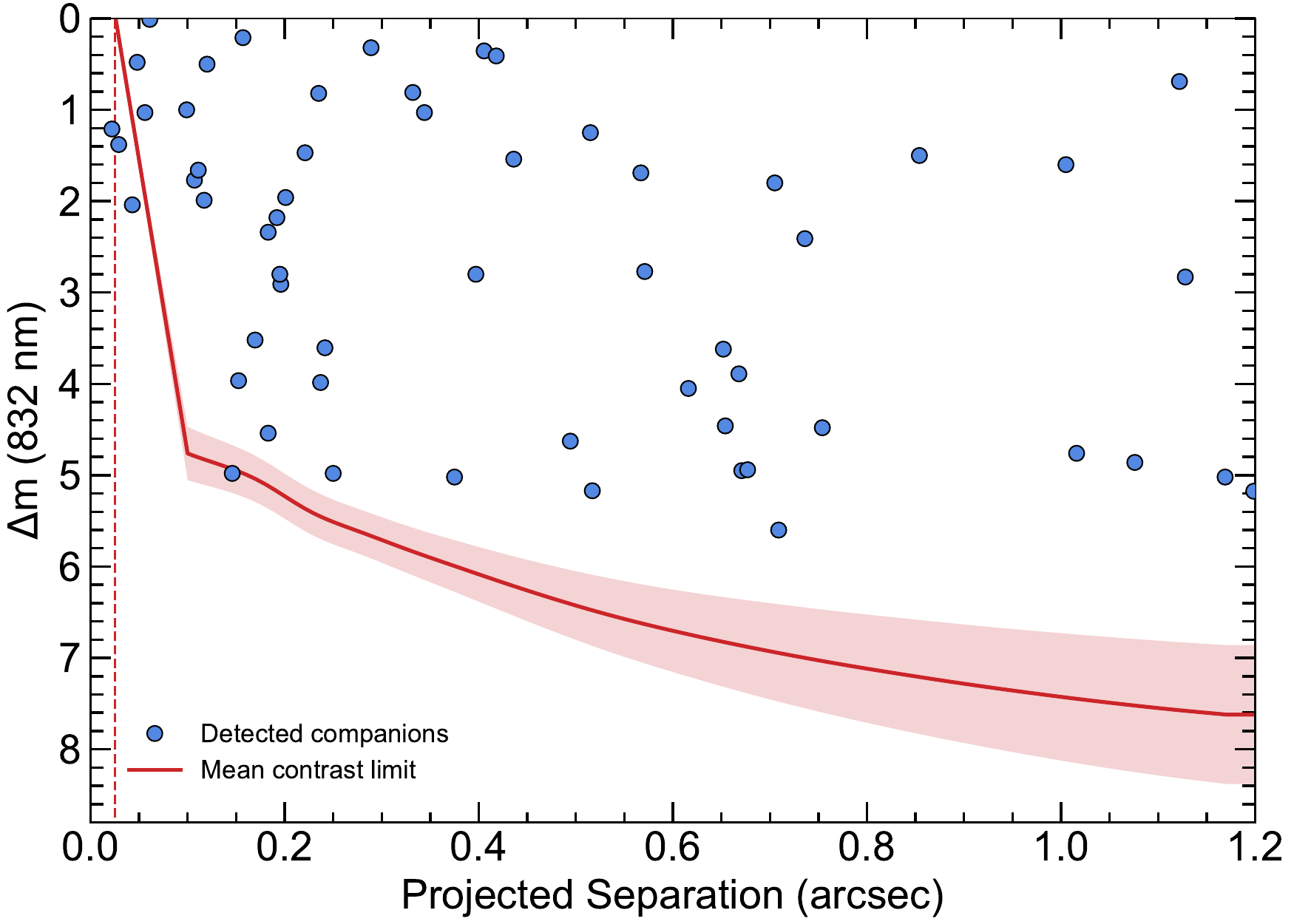}
\includegraphics[width=0.49\textwidth]{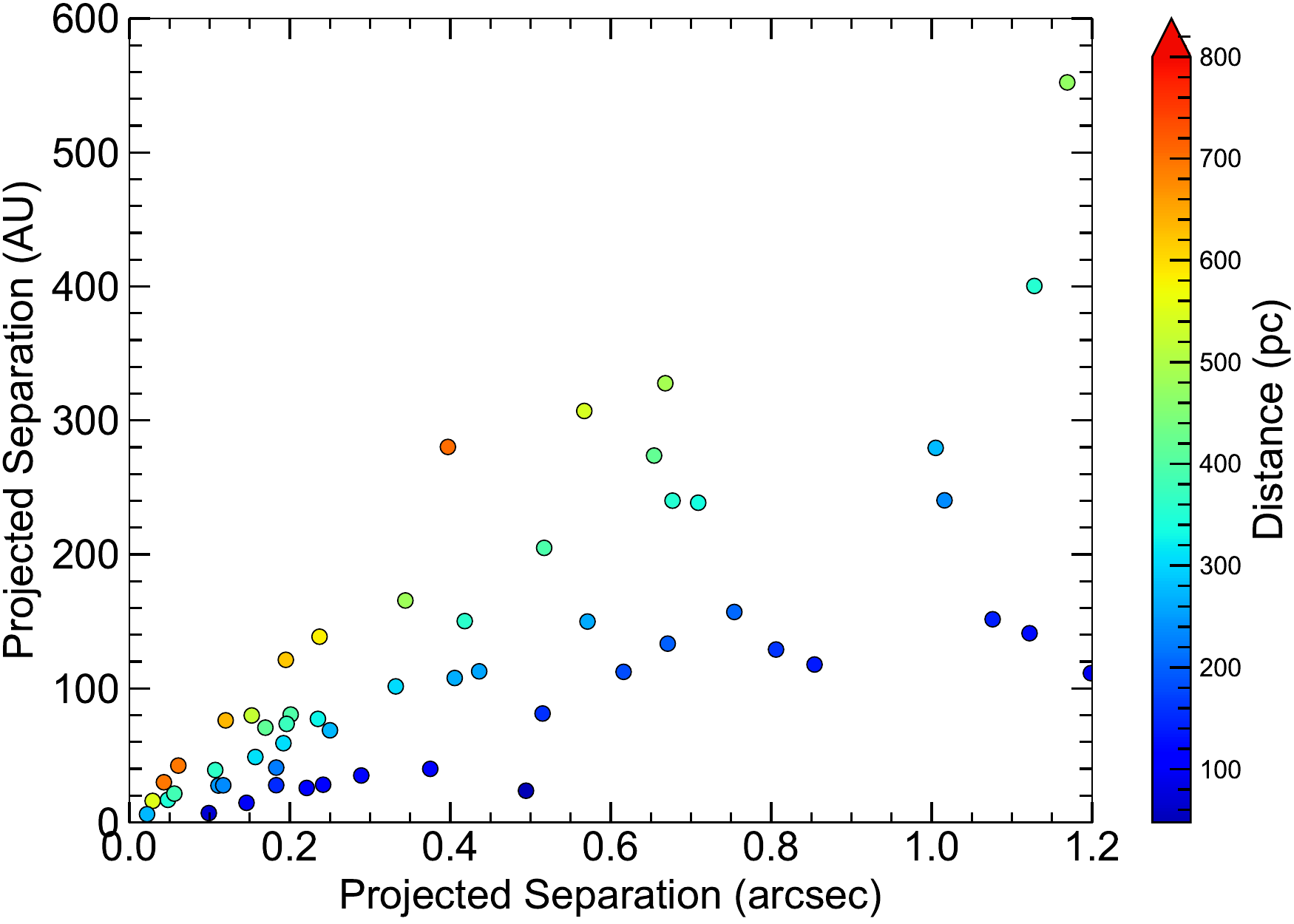}
\caption{Properties of the \numcompexo companions with separations less than 1.2" around exoplanet host stars. \textit{Left:} The magnitude difference as a function of the angular separation. The detected companions are shown as the solid blue points. The average $5\sigma$ detection limit is shown as the solid red line, and the shading corresponds to the typical spread in this limit. The few companions below the contrast limit correspond to observations taken during better seeing or at higher signal-to-noise. The Gemini angular resolution limit in the 832~nm filter is shown as the vertical red line. \textit{Right:} Comparison of the observed angular separation of our detected companions and the derived spatial separation determined using Gaia distances. The color corresponds to the system distance. 
\label{companions}}
\end{figure*}

\begin{figure*}
\centering
\includegraphics[width=\textwidth]{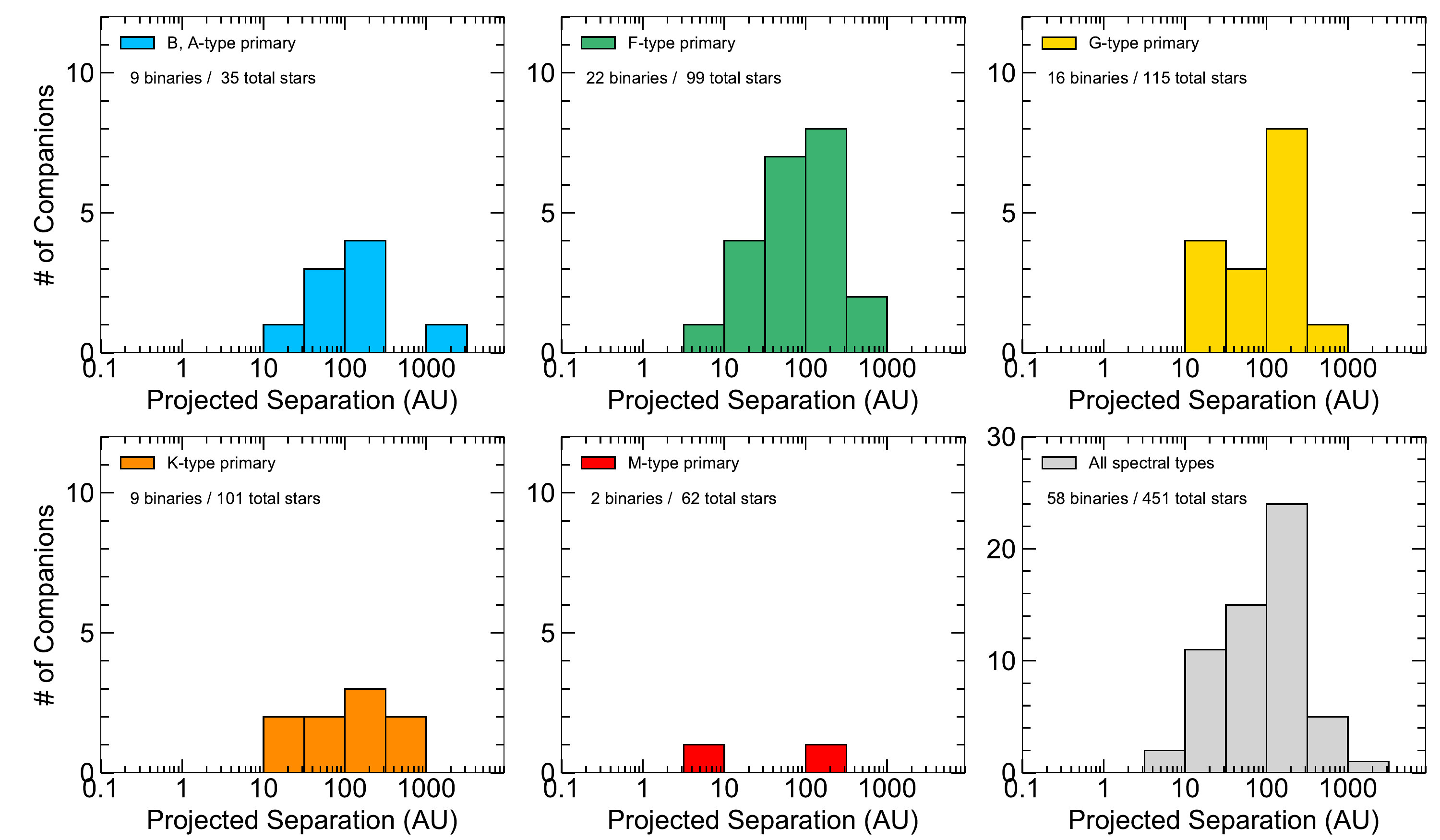}
\caption{Histograms of the projected physical separation (i.e., semi-major axis) of the exoplanet host binaries in our sample on a logarithmic scale and broken down by the spectral type of the primary star. 
\label{companionteff}}
\vspace{0.5cm}
\end{figure*}

Next, we calculated the separation in AU from the angular separation and the Gaia distance for each system, shown in the right panel of Figure~\ref{companions}. We then estimated the binary orbital period using Kepler's Third Law, the primary component's mass estimated from the effective temperature, the secondary component's mass calculated from the mass ratio, and the semi-major axis equal to the instantaneous separation measured with our speckle observations. This last assumption is discussed further in Section~\ref{sect-sim-ratio}. The estimated semi-major axes and orbital periods are listed in the last columns of Table~\ref{companiontable}.  Figure~\ref{companionteff} shows the logarithmic distribution of companion separation broken down by spectral type of the primary star. The distributions look similar for all spectral types, except the M-stars for which we only detected two companions. This is likely due to their decreased multiplicity rate (27\%) and the smaller peak separation (20~AU) of M-type field binaries \citep{winters19}, so most close companions could not be resolved by speckle interferometry. These stars are also quite faint, so our observations may not have achieved sufficiently deep contrast limits to detect fainter companions.
For our comparison with different model predictions, we only consider binaries with an F-, G-, or K-type primary (\numcompexofgk systems), which make up the majority of our sample.

\clearpage

\break

\startlongtable
\begin{deluxetable*}{lccccccccc}
\tablewidth{0pt}
\tablecolumns{10}
\tablecaption{Companions Detected \label{companiontable}    }
\tablehead{
\colhead{Target} 
& \colhead{\ \ $\rho$ (\arcsec) \ \ } & \colhead{\ \ $\theta$ (deg)\ \ } & \colhead{$\Delta$m (mag)} 
& \colhead{\ \ $\rho$ (\arcsec) \ \ } & \colhead{\ \ $\theta$ (deg)\ \ } & \colhead{$\Delta$m (mag)} 
& \colhead{$M_2/M_1$} & \colhead{$a$ (AU)} & \colhead{Binary}  
\\  
\colhead{} 
& \colhead{(562 nm)} & \colhead{(562 nm)} & \colhead{(562 nm)}  
& \colhead{(832 nm)} & \colhead{(832 nm)} & \colhead{(832 nm)}  
& \colhead{ } & \colhead{ } & \colhead{Period (yr)}  
}
\startdata 
 & \multicolumn{7}{c}{Exoplanet Host Stars} & \\  
TOI  140 &   0.044 &   336.2 &    0.76 &   0.048 &   337.0 &  0.48 &   0.88 &   16.8 &    46  \\
TOI  141$^*$ & \nodata & \nodata & \nodata &   0.494 &   240.6 &  4.63 &   0.47 &   23.6 &    93  \\
         & \nodata & \nodata & \nodata &   1.333 &   306.8 &  5.60 &   0.38 &   63.7 &   430  \\
TOI  172 &   1.175 &   319.3 &    5.20 &   1.262 &   321.9 &  5.36 &   0.43 &  452.9 &  7900  \\
TOI  224 &   0.100 &   224.8 &    2.07 &   0.099 &   226.6 &  1.00 &   0.72 &    6.9 &    20  \\
TOI  235 &   0.850 &   289.2 &    1.78 &   0.854 &   291.4 &  1.50 &   0.75 &  117.8 &  1000  \\
TOI  245 &   0.990 &   109.1 &    3.54 &   1.206 &   109.6 &  2.39 &   0.59 &  148.3 &  1300  \\
TOI  252 &   0.436 &   307.8 &    2.16 &   0.436 &   309.9 &  1.54 &   0.79 &  112.7 &  1000  \\
TOI  264 &   0.655 &   329.5 &    5.38 &   0.654 &   331.7 &  4.46 &   0.45 &  273.8 &  3400  \\
TOI  271 & \nodata & \nodata & \nodata &   0.146 &   231.1 &  4.98 &   0.43 &   14.6 &    44  \\
TOI  287 & \nodata & \nodata & \nodata &   0.152 &   125.9 &  3.96 &   0.52 &   79.8 &   570  \\
TOI  291 &   0.035 &   146.2 &    0.65 &   0.029 &   145.2 &  1.38 &   0.82 &   16.0 &    46  \\
TOI  309 &   0.330 &    77.6 &    1.27 &   0.344 &    77.7 &  1.03 &   0.81 &  165.6 &  1700  \\
TOI  325 & \nodata & \nodata & \nodata &   0.616 &   221.8 &  4.05 &   0.49 &  112.3 &  1200  \\
TOI  354 &   0.054 &    87.1 &    1.08 &   0.056 &    87.8 &  1.03 &   0.82 &   21.5 &    73  \\
TOI  364 &   0.384 &    95.2 &    0.36 &   0.405 &    96.6 &  0.35 &   0.93 &  107.8 &   730  \\
TOI  380 & \nodata & \nodata & \nodata &   0.201 &     6.1 &  1.96 &   0.73 &   80.4 &   590  \\
TOI  402 & \nodata & \nodata & \nodata &   1.467 &   234.2 &  9.33 &   0.14 &   65.8 &   540  \\
TOI  457 & \nodata & \nodata & \nodata &   1.122 &   223.8 &  0.69 &   0.83 &  141.2 &  3000  \\
TOI  487 &   0.515 &   198.7 &    1.53 &   0.515 &   200.8 &  1.25 &   0.77 &   81.2 &   570  \\
TOI  492 & \nodata & \nodata & \nodata &   0.668 &   321.1 &  3.89 &   0.49 &  327.7 &  4400  \\
TOI  564 & \nodata & \nodata & \nodata &   0.671 &    73.9 &  4.95 &   0.44 &  133.4 &  1300  \\
TOI  568 &   0.310 &   189.8 &    0.90 &   0.332 &   192.0 &  0.81 &   0.85 &  101.5 &   740  \\
TOI  573 &   1.442 &   227.6 &    1.30 &   1.600 &   234.3 &  1.25 &   0.69 &  142.0 &  2200  \\
TOI  676 & \nodata & \nodata & \nodata &   1.331 &   103.8 &  2.20 &   0.68 &  944.1 & 23300  \\
TOI  680 & \nodata & \nodata & \nodata &   0.806 &   331.4 &  \nodata & \nodata &  129.0 &  1400  \\
TOI  697$^*$ & \nodata & \nodata & \nodata &   1.199 &   137.9 &  5.17 &   0.41 &  111.4 &  1000  \\
         & \nodata & \nodata & \nodata &   1.269 &   139.4 &  5.46 &   0.38 &  117.9 &  1100  \\
TOI  722 &   0.224 &   114.1 &    2.46 &   0.221 &   115.8 &  1.47 &   0.72 &   25.7 &   120  \\
TOI  791 & \nodata & \nodata & \nodata &   0.709 &    25.6 &  5.60 &   0.37 &  238.6 &  2700  \\
TOI  798 &   0.289 &   179.4 &    0.64 &   0.289 &   181.5 &  0.32 &   0.93 &   35.0 &   180  \\
TOI  854 &   0.023 &   204.6 &    1.35 &   0.022 &   211.1 &  1.21 &   0.79 &    6.1 &    11  \\
TOI  906 &   1.294 &    49.0 &    3.67 &   1.375 &    50.7 &  2.98 &   0.57 &  186.2 &  1900  \\
TOI  926 &   0.192 &   156.9 &    2.53 &   0.192 &   158.8 &  2.18 &   0.68 &   59.1 &   350  \\
TOI  931 &   0.113 &   234.9 &    0.65 &   0.120 &   237.6 &  0.50 &   0.90 &   76.1 &   460  \\
TOI  952$^*$ & \nodata & \nodata & \nodata &   0.169 &    72.1 &  3.52 &   0.48 &   70.7 &   390  \\
         & \nodata & \nodata & \nodata &   1.262 &   135.4 &  4.91 &   0.39 &  526.3 &  8300  \\
TOI 1032 &   1.161 &    81.6 &    2.16 &   1.329 &    83.0 &  1.61 &   0.59 &  796.0 & 11300  \\
TOI 1035 &   0.101 &   100.2 &    1.96 &   0.107 &   112.1 &  1.77 &   0.71 &   39.1 &   180  \\
TOI 1037 & \nodata & \nodata & \nodata &   0.250 &   242.1 &  4.98 &   0.32 &   68.7 &   340  \\
TOI 1039 &   0.708 &   285.6 &    4.02 &   0.736 &   286.3 &  2.41 &   0.38 & 1121.6 & 15500  \\
TOI 1131 &   0.095 &   214.4 &    1.45 &   0.111 &   214.0 &  1.66 &   0.75 &   27.5 &   110  \\
TOI 1133 &   0.550 &   102.1 &    3.05 &   0.571 &   102.4 &  2.77 &   0.59 &  149.9 &  1300  \\
TOI 1152 &   1.074 &    22.6 &    0.93 &   1.284 &    22.7 &  0.85 &   0.83 &  133.9 &  1200  \\
TOI 1189 &   0.984 &   318.5 &    1.36 &   1.005 &   319.1 &  1.60 &   0.78 &  279.5 &  3700  \\
TOI 1191 & \nodata & \nodata & \nodata &   0.677 &    85.8 &  4.94 &   0.40 &  240.1 &  2600  \\
TOI 1192 &   0.187 &   200.6 &    3.37 &   0.196 &   201.4 &  2.91 &   0.55 &   73.5 &   440  \\
TOI 1204 & \nodata & \nodata & \nodata &   0.375 &    38.8 &  5.02 &   0.39 &   39.9 &   180  \\
TOI 1217 &   1.088 &   242.7 &    5.02 &   1.169 &   242.7 &  5.02 &   0.44 &  552.3 &  9700  \\
TOI 1261 & \nodata & \nodata & \nodata &   1.722 &   316.9 &  2.87 &   0.59 &  344.2 &  4900  \\
TOI 1293 & \nodata & \nodata & \nodata &   1.016 &   120.4 &  4.76 &   0.45 &  240.4 &  3000  \\
TOI 1335 &   0.057 &   303.1 &    0.01 &   0.061 &   304.1 &  0.01 &   0.99 &   42.4 &   150  \\
TOI 1343 & \nodata & \nodata & \nodata &   0.517 &   333.6 &  5.17 &   0.30 &  204.9 &  1700  \\
TOI 1440 & \nodata & \nodata & \nodata &   1.536 &   324.9 &  \nodata &  \nodata &  363.2 &  6800  \\
TOI 1447 &   0.150 &    67.7 &    0.22 &   0.157 &    67.7 &  0.21 &   0.96 &   48.8 &   200  \\
TOI 1491 & \nodata & \nodata & \nodata &   0.754 &   269.5 &  4.48 &   0.47 &  157.0 &  1600  \\
TOI 1531 &   0.565 &   199.2 &    1.11 &   0.567 &   199.6 &  1.69 &   0.82 &  307.1 &  4500  \\
TOI 1544 &   0.624 &   129.3 &    3.35 &   0.652 &   129.9 &  3.62 &   0.58 &  641.6 & 15200  \\
TOI 1559 &   0.039 &   122.6 &    2.08 &   0.043 &   123.0 &  2.04 &   0.61 &   29.9 &    92  \\
TOI 1560 &   1.246 &    62.7 &    4.03 &   1.366 &    68.4 &  3.08 &   0.44 &  731.2 & 11400  \\
TOI 1565 &   0.233 &   207.3 &    0.85 &   0.235 &   207.7 &  0.82 &   0.81 &   77.2 &   360  \\
TOI 1598 & \nodata & \nodata & \nodata &   1.076 &   304.3 &  4.86 &   0.44 &  151.6 &  1500  \\
TOI 1636 & \nodata & \nodata & \nodata &   0.183 &   356.3 &  2.34 &   0.72 &   27.7 &   130  \\
TOI 1641 &   0.194 &   296.1 &    3.53 &   0.195 &   296.4 &  2.80 &   0.46 &  121.3 &   730  \\
TOI 1677 & \nodata & \nodata & \nodata &   0.183 &   142.0 &  4.54 &   0.47 &   40.9 &   210  \\
TOI 1678 &   0.336 &   307.1 &    4.01 &   0.397 &   307.5 &  2.80 &   0.49 &  280.2 &  2800  \\
TOI 1719 &   0.116 &   193.3 &    1.78 &   0.117 &   193.8 &  1.99 &   0.70 &   27.7 &   100  \\
TOI 1740 & \nodata & \nodata & \nodata &   0.241 &   258.6 &  3.60 &   0.56 &   28.1 &   130  \\
TOI 2010 & \nodata & \nodata & \nodata &   1.578 &   138.0 &  \nodata &  \nodata &  171.1 &  2200  \\
TOI 2035 &   0.415 &   245.9 &    0.45 &   0.418 &   246.2 &  0.41 &   0.87 &  150.2 &   900  \\
TOI 2220 & \nodata & \nodata & \nodata &   1.128 &   251.2 &  2.83 &   0.61 &  400.3 &  6200  \\
TOI 2305 & \nodata & \nodata & \nodata &   0.237 &   203.1 &  3.98 &   0.52 &  138.6 &  1300  \\
TOI 2310 &   0.708 &   332.5 &    2.81 &   0.705 &   334.8 &  1.80 &   0.71 &  650.9 & 13600  \\[6pt]
\hline 
\multicolumn{9}{c}{Transit False Positive Stars} \\ 
TOI  138 &   0.099 &   258.9 &    1.38 &   0.108 &   260.5 &  1.33 &   0.78 &   20.7 &    94  \\
TOI  146 &   0.302 &    75.0 &    1.15 &   0.324 &    77.0 &  1.15 &   0.80 &  161.8 &  2100  \\
TOI  154 & \nodata & \nodata & \nodata &   0.154 &   202.8 &  1.28 &   0.77 &   61.8 &   470  \\
TOI  311 &   0.039 &   130.6 &    0.91 &   0.028 &   148.8 &  0.96 &   0.85 &    6.5 &    21  \\
TOI  372 & \nodata & \nodata & \nodata &   0.267 &   262.3 &  2.45 &   0.64 &  137.8 &  1600  \\
TOI  378 &   0.098 &   167.9 &    1.29 &   0.137 &   182.0 &  1.49 &   0.77 &  316.2 &  5400  \\
TOI  379 &   0.053 &    19.5 &    0.07 &   0.052 &    21.3 &  0.41 &   0.95 &   12.0 &    41  \\
TOI  463 & \nodata & \nodata & \nodata &   0.816 &   214.4 &  3.78 &   0.48 &  150.3 &  1600  \\
TOI  593 & \nodata & \nodata & \nodata &   0.664 &   170.1 &  3.81 &   0.49 &  255.5 &  3700  \\
TOI  884 &   0.049 &   323.2 &    1.79 &   0.052 &   324.1 &  1.69 &   0.56 &   52.7 &   210  \\
TOI  894 &   0.619 &   329.5 &    3.32 &   0.649 &   331.0 &  2.41 &   0.49 &  233.1 &  2300  \\
TOI  896 & \nodata & \nodata & \nodata &   0.062 &   126.2 &  3.48 &   0.49 &    9.6 &    25  \\
TOI  898 &   1.306 &    34.9 &    2.11 &   1.358 &    35.4 &  2.07 &   0.70 &  651.2 & 16800  \\
TOI  919 &   0.088 &    83.7 &    0.71 &   0.095 &    85.9 &  0.48 &   0.89 &   31.4 &   170  \\
TOI  935 & \nodata & \nodata & \nodata &   0.785 &   146.5 &  2.19 &   0.74 &  296.7 &  5900  \\
TOI  995 &   0.116 &   137.9 &    2.22 &   0.121 &   139.5 &  2.54 &   0.73 &  151.1 &  2100  \\
TOI  996 &   0.143 &   228.7 &    1.99 &   0.153 &   230.9 &  1.87 &   0.64 &   79.1 &   520  \\
TOI 1038 &   0.443 &    99.9 &    1.84 &   0.474 &   102.1 &  1.42 &   0.66 &  119.5 &   910  \\
TOI 1047 &   0.826 &   246.4 &    2.83 &   0.883 &   248.7 &  2.31 &   0.66 &  119.5 &  1300  \\
TOI 1051 &   0.368 &   241.0 &    1.12 &   0.368 &   243.1 &  1.00 &   0.81 &   23.2 &   110  \\
TOI 1093 &   0.102 &   108.2 &    2.21 &   0.109 &   110.7 &  2.04 &   0.70 &   64.6 &   680  \\
TOI 1147 & \nodata & \nodata & \nodata &   0.377 &   346.0 &  3.62 &   0.51 &   77.9 &   620  \\
TOI 1193 &   0.087 &   255.3 &    0.62 &   0.090 &   255.4 &  0.59 &   0.88 &   77.9 &   550  \\
TOI 1417 &   0.031 &   307.0 &    2.71 &   0.027 &   307.2 &  2.63 &   0.57 &    4.1 &     6  \\
TOI 1487 &   0.097 &    45.6 &    4.20 & \nodata & \nodata &  \nodata& 0.55 &   10.6 &    39  \\
TOI 1543 & \nodata & \nodata & \nodata &   0.320 &   178.5 &  5.02 &   0.32 &  194.4 &  1800  \\
TOI 1557 & \nodata & \nodata & \nodata &   1.406 &   135.7 &  5.59 &   0.38 &  407.3 &  8200  \\
TOI 1662 &   0.226 &   347.4 &    2.52 &   0.227 &   347.5 &  2.11 &   0.61 &   79.3 &   560  \\
TOI 1673 &   0.274 &    50.5 &    1.61 &   0.276 &    50.9 &  1.50 &   0.75 &   58.2 &   450  \\

\enddata  
\tablenotetext{*}{Found to be a triple system, so the parameters for each companion are listed separately.}
\end{deluxetable*}

\begin{figure*}
\centering
\includegraphics[width=\textwidth]{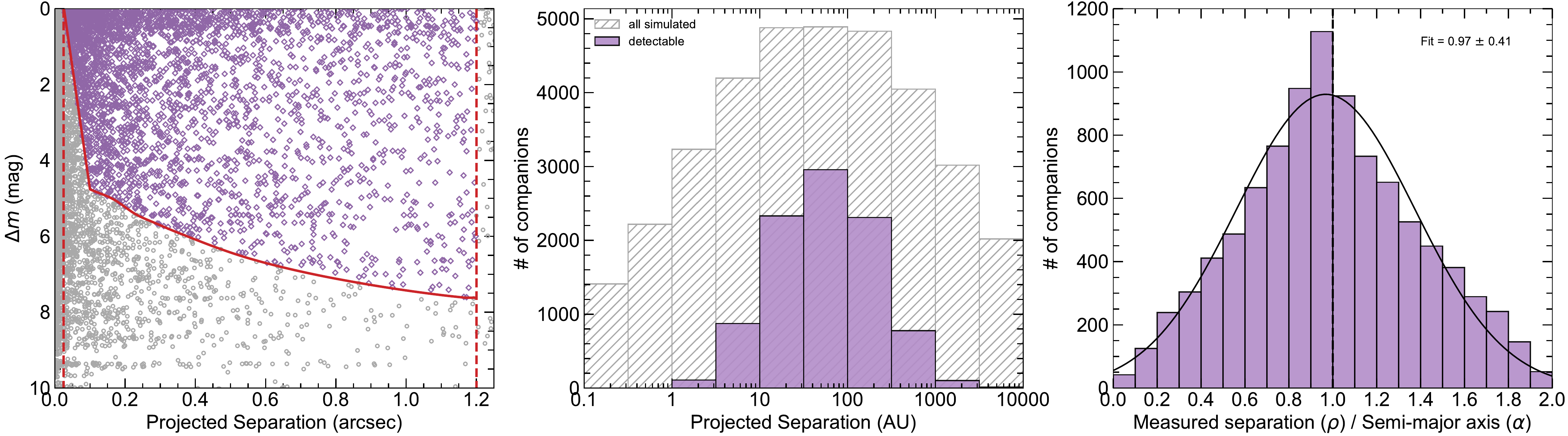}
\caption{Results of our simulated binary sample based on orbital demographics of field binaries from \citet{raghavan10}. 
\textit{Left:} Magnitude difference as a function of the angular separation for a subset of the simulated companions. The detectable companions are shown as purple diamonds, while the undetectable companions are shown as grey circles. The mean Gemini detection limit is shown as the solid red line, and the angular resolution limits are shown as the dashed red lines. 
\textit{Middle:} Logarithmic distribution of the projected physical separation for the full simulated binary sample. The filled, purple histograms represent the detectable companions, while the grey, hatched histograms represent the full sample. 
\textit{Right:} Histogram of the ratio between the instantaneous separation and the modeled semi-major axis of the detectable companions. The gaussian fit is shown in solid black, and a dashed line at unity is shown for reference. 
\label{sim_sample}}
\end{figure*}

\section{Simulated Binaries}\label{sect-sim}

\subsection{Field Binary Model}
We simulated a series of binary stars following the methods in \citet{matson18} and \citet{ziegler20} in order to compare our results to the predictions for field binaries and investigate the suppression of close companions around exoplanet hosts. We began by selecting only F-, G-, and K-type exoplanet candidate hosts from our Gemini TOI sample (337 stars) and the sample observed with WIYN by \citet[][130 stars]{howell21}, then used these masses and distances as the inputs for our simulations. For each of a thousand iterations, we randomly chose 46\% of these stars to be the primary components of our simulated binaries following the multiplicity rate of Solar-type field stars found by \citet{raghavan10}. Masses for the secondary components were then drawn randomly from the distribution in \citet{raghavan10}: uniform probability for $M_2/M_1 = 0.4-0.9$ then a higher probability for $0.9-1.0$. The absolute $I$-band magnitude of each component was interpolated based on mass using the Modern Mean Dwarf Sequence \citep{pecaut13} in order to calculate a magnitude difference for each system.  

Next, orbital parameters were randomly assigned to each system from the log-normal period distribution of \citet[$\log P =5.03$ days, $\sigma=2.28$]{raghavan10}, the eccentricity distribution of \citet{dm91}, and uniform distributions for the inclination, angular semi-major axis ($\alpha$), longitude of periastron, and longitude of the ascending node. Finally, we calculated the projected angular separation ($\rho$) of each companion at a random orbital phase to mimic our speckle observations, then identified ``detectable" companions that lie within the Gemini angular resolution (inner) limit of 0.025" and the speckle correlation (outer) limit of 1.2" and have a magnitude difference within the mean 832~nm Gemini contrast limit. 

The results of our simulated binary sample are shown in Figure~\ref{sim_sample}. The left panel shows the companion magnitude difference versus separation for several iterations of our simulation, with detectable companions plotted in purple and the undetectable companions plotted in grey. The middle panel shows a histogram of the physical separation on a logarithmic scale for all simulated companions. We averaged the results from each iteration to predict the number of companions detectable with speckle imaging in each 0.5~dex bin, and estimated the uncertainty from the standard deviation across all iterations. This expected number of companions is compared to the observed number in the next section (Section~\ref{results2}). We note that WIYN has a slightly larger angular resolution limit and shallower contrast limit than Gemini, and therefore is less sensitive to close companions, but this difference between telescopes is within the simulation's uncertainties. Lastly, the right panel of Figure~\ref{sim_sample} shows the ratio between the instantaneous separation and the modeled semi-major axis of the detectable companions, which we discuss in Section~\ref{sect-sim-ratio}.

\begin{figure*}
\centering
\includegraphics[width=\textwidth]{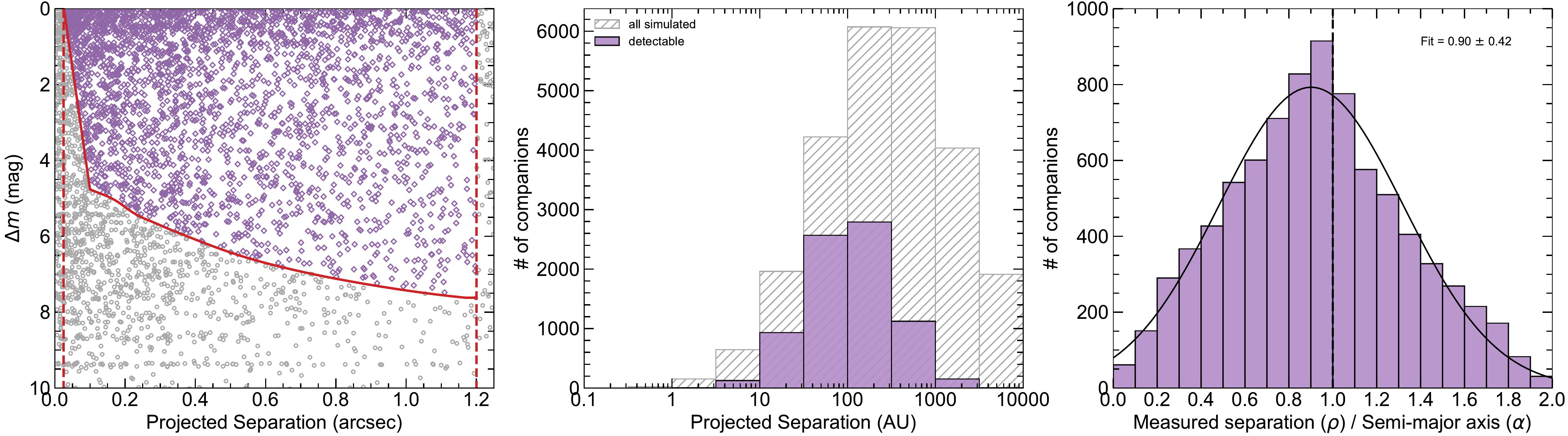}
\caption{Same as Figure~\ref{sim_sample}, but for simulated binaries created using a log-normal period distribution with the peak at $\log P=6.2$ days and width of $\sigma=1.2$.
\label{sim_long}}
\end{figure*}

\begin{figure*}
\centering
\includegraphics[width=\textwidth]{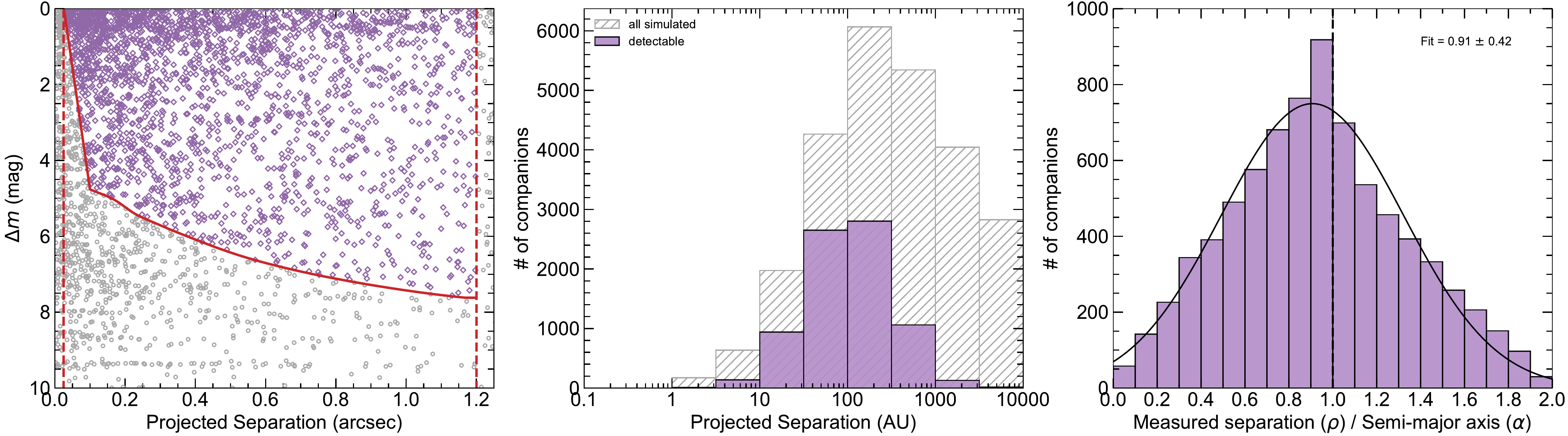}
\caption{Same as Figure~\ref{sim_sample}, but for simulated binaries created using the \citet{moe19} model. 
\label{sim_sup}}
\end{figure*}

\subsection{Additional Period Distributions}
Recent work has found that exoplanet host binaries have larger separations than field binaries; for example, \citet{howell21} suggests a peak at $\log P = 5.5$ days. We used a grid search to test several different log-normal period distributions against the distribution of companions seen by Gemini and WIYN. We found that a log-normal distribution with a peak at $\log P = 6.2$ days and width of $\sigma=1.2$ minimized the $\chi^2$ between the simulated and observed companion separations. The simulation results using this ``longer period" model are shown in Figure~\ref{sim_long}. This distribution agrees with that found by \citet{howell21}. 

Lastly, we created simulated binaries based on the model of \citet{moe19} representing the lack of close binaries as planet hosting systems. They calculated the ratio between the number of exoplanet host binaries and the number of field binaries as a function of the binary separation. For this model, we started with the \citet{raghavan10} period distribution, then removed a fraction of close companions within each separation range used by \citet{moe19}. Within 1~AU, all companions are removed. From $1-200$~AU, the survival fraction increases linearly in log-space. All companions outside 200~AU are kept. These simulation results are shown in Figure~\ref{sim_sup} and have a median period of $\log P = 6.4$ days similar to the longer period model, but include more stars at large separations (typically past 1.2" where speckle interferometry is not sensitive).  

\begin{figure*}
\centering
\includegraphics[width=1.02\textwidth]{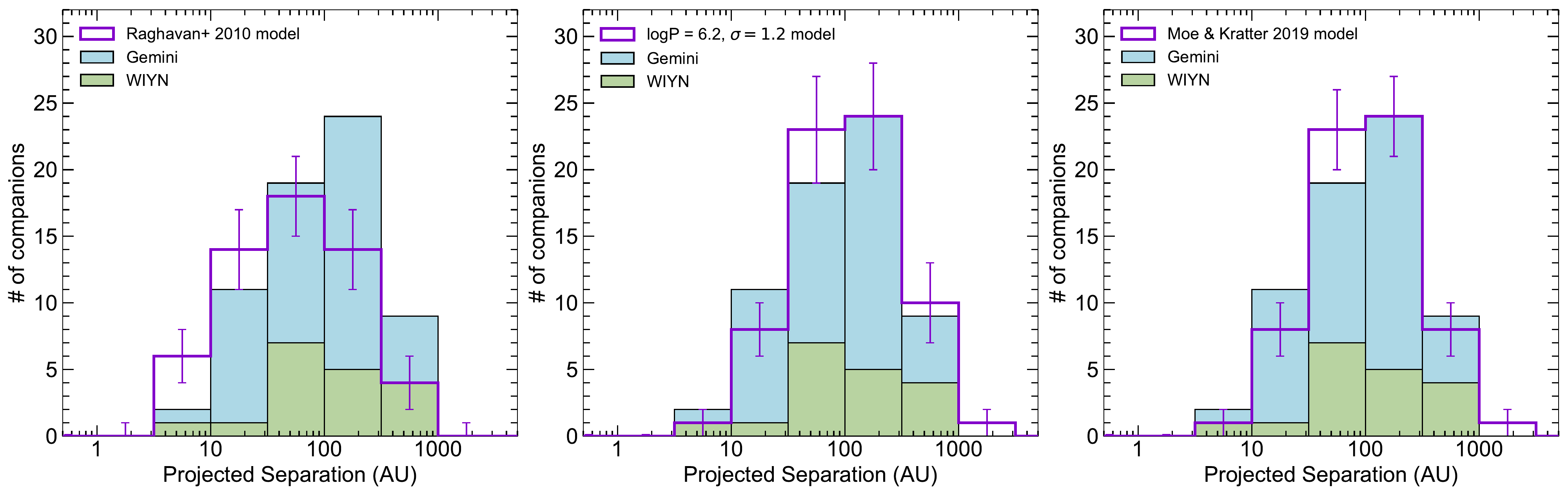}
\caption{Histograms for the projected physical separation in logarithmic bins. The solid bars correspond to the companions with separations less than 1.2" around solar-type exoplanet host stars detected by Gemini (blue) and WIYN \citep[green]{howell21}. The purple unfilled histograms correspond to the expected number of companions based on the field binary period distribution (left), on a distribution peaking at longer periods (middle), and on the \citet{moe19} model (right). 
\label{logsepau}}
\end{figure*}

\subsection{Testing Binary Separations}\label{sect-sim-ratio}
When we observe a binary at a random orbital phase, are we more likely to under-estimate or over-estimate the separation compared to the true semi-major axis? To answer this question, we compared the projected angular separation to the assigned semi-major axis for all detectable companions. The right panels of Figure~\ref{sim_sample}-\ref{sim_sup} show a histogram of the ratio between these two parameters ($\rho / \alpha$). We then fit a Gaussian to this distribution and found that the mean ratio is $0.97 \pm 0.41$ for the \citet{raghavan10} model, $0.90 \pm 0.42$ for the longer period model, and $0.91\pm0.42$ for the \citet{moe19} model. These ratio distributions are statistically consistent based on p-values of 0.05 found from comparing each distribution to the others using a Kolmogorov-Smirnov test. Therefore, our observed separation distribution is only slightly underestimated, and the projected separation measured in one observation is likely a reasonable estimate until the true semi-major axis can be found by resolving the full visual orbit.

\section{Results}\label{sect-results}
\newcommand{\numgemwiyn}{65 }  
\newcommand{\numexpsam}{56 }    
\newcommand{\numexplong}{67 }   
\newcommand{\numexpsup}{65 }    

\subsection{Binary Exoplanet Host Orbital Period Distribution}\label{results2}
Figure~\ref{logsepau} shows the logarithmic separation distribution for the \numgemwiyn companions of Solar-type exoplanet hosts found by speckle imaging at Gemini (this work) and WIYN \citep{howell21}, compared to the predictions of three different binary period distributions. All predictions use a companion frequency of 46\%. The observed companions peak at larger separations ($\sim$100 AU) than predicted by the field binary model \citep[50 AU,][]{raghavan10}. The field binary model has a reduced $\chi^2$ value of 4.5 and predicted a total of \numexpsam companions, indicating that this period distribution is also too wide. Next, the ``longer period model" ($\log P = 6.2$ days, $\sigma=1.2$) is based on the observed distribution, so naturally peaks at larger separations. This model has the lowest reduced $\chi^2$ value of 1.1 and predicts \numexplong companions. Lastly, the \citet{moe19} model matches the observed distribution well with a reduced $\chi^2$ value of 1.3. It also best matches the total number of companions by predicting \numexpsup would be detected. Overall, both the longer period model and the \citet{moe19} model provide good fits to the observed companion distribution, but we can not statistically distinguish between the two. Future work to include common proper motion binaries from Gaia could  help differentiate between these models at larger orbital separations \citep[e.g.,][]{ziegler21}. Lastly, excluding the APC systems from our sample would remove up to two companions from each histogram bin, but the overall shape would remain intact. The observed companions would still peak at 100~AU, supporting our conclusion that binary exoplanet hosts have larger separations than field binaries.

\subsection{Exoplanet Candidate Properties\label{results3}}
We investigated if any correlations existed between the exoplanet candidate properties (such as radius or orbital period), the presence of a stellar companion, and the orbital properties of that companion. The planet parameters were taken from ExoFOP\footnote{The planet radii were estimated using the stellar radii from the TESS Input Catalog, which does not account for the multiplicity of the host stars and therefore could underestimate the radii of stars with bright, unresolved companions.}, and the radii of planets in binary systems have been corrected for ``third-light" contamination from the companion star that dilutes the transits \citep{ciardi15}. Table~\ref{planettable} lists the radius correction factor, apparent radius from ExoFOP (when available), and the corrected radius for each planet found in a binary system with $\rho<1.2"$, assuming the planet orbits the brighter, primary component. Our sample includes some planet candidates with radii larger than $20 R_\oplus$, which are likely red dwarf companions.

\begin{figure*}
\centering
\includegraphics[width=\textwidth]{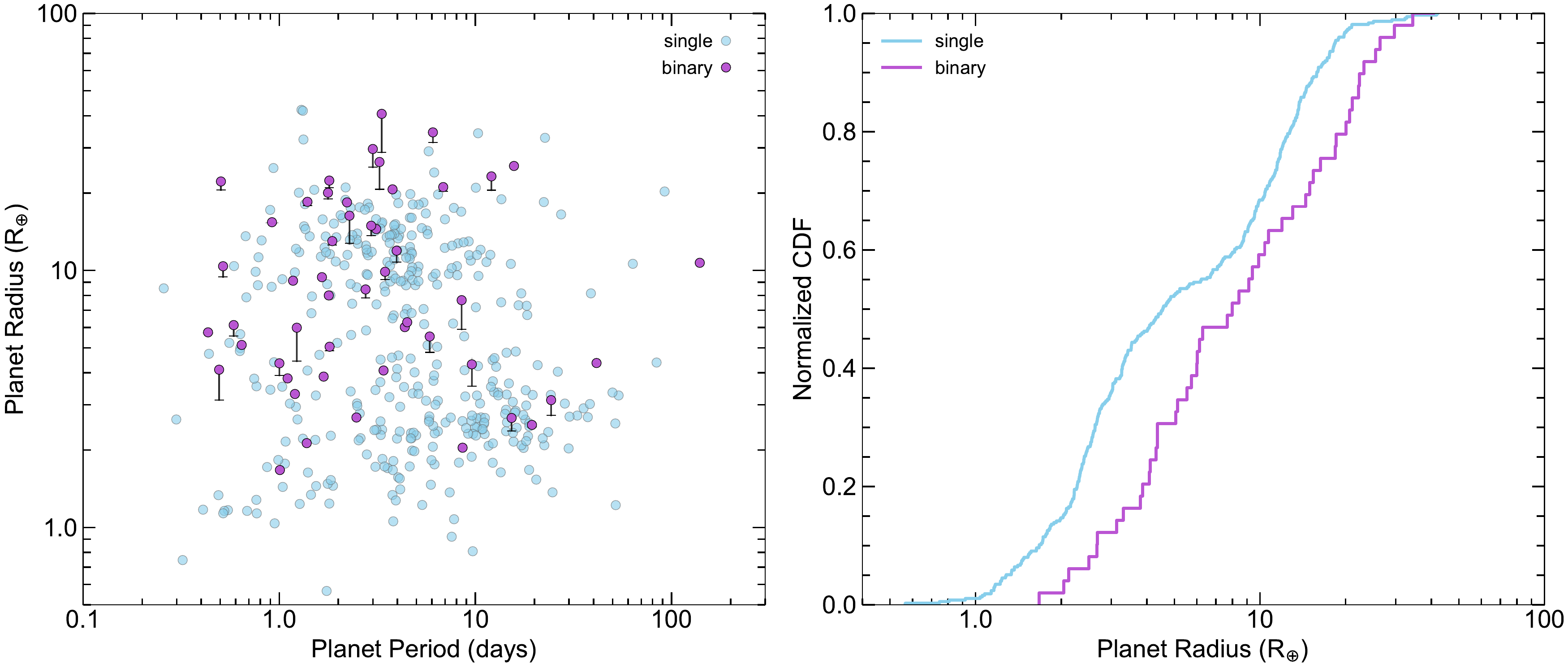}
\caption{\textit{Left:} Planet radius as a function of orbital period for planet candidate hosts in single star systems (light blue) and in close binary systems (purple). For the planets in binaries, we show the increase in radius (tail to plotted point) when correcting for dilution by the companion. \textit{Right:} Normalized cumulative distribution function for the radii of planets in single and binary systems. The two subsamples appear similar, except for the clear observational bias against detecting Earth-sized, transiting planets in binary systems. 
\label{single_binary}}
\end{figure*}

The left panel of Figure~\ref{single_binary} shows the planet radius as a function of the planet's period, color coded by whether the host star is single or part of a binary system. Giant planet candidates are seen at all periods independant of host binarity. However, we see an observational bias against the detection of Earth-sized ($R_{pl}< 2~R_\oplus$), transiting planet candidates in binary star systems. Using the Kolmogorov-Smirnov test to compare the radius distribution of planets in binaries to those around single stars, we found a p-value of 0.001 indicating that these subsamples are statistically different. This is also evident in the right panel of Figure~\ref{single_binary} that shows the cumulative distribution function for these two subsamples and the lack of Earth-sized, transiting planets in binary systems. \citet{saval20} investigated this observational bias and found that the occurence rates for super-Earths and sub-Neptunes are underestimated by roughly 26\% if transit dilution from stellar companions is not taken into account. 
In addition, this effect will manifest itself more strongly as the planet orbital period becomes longer (fewer measured transits, lower signal-to-noise ratio) making planets approaching or in the Habitable Zone particularly susceptible to this bias. 

\begin{figure*}
\centering
\includegraphics[height=0.88\textheight]{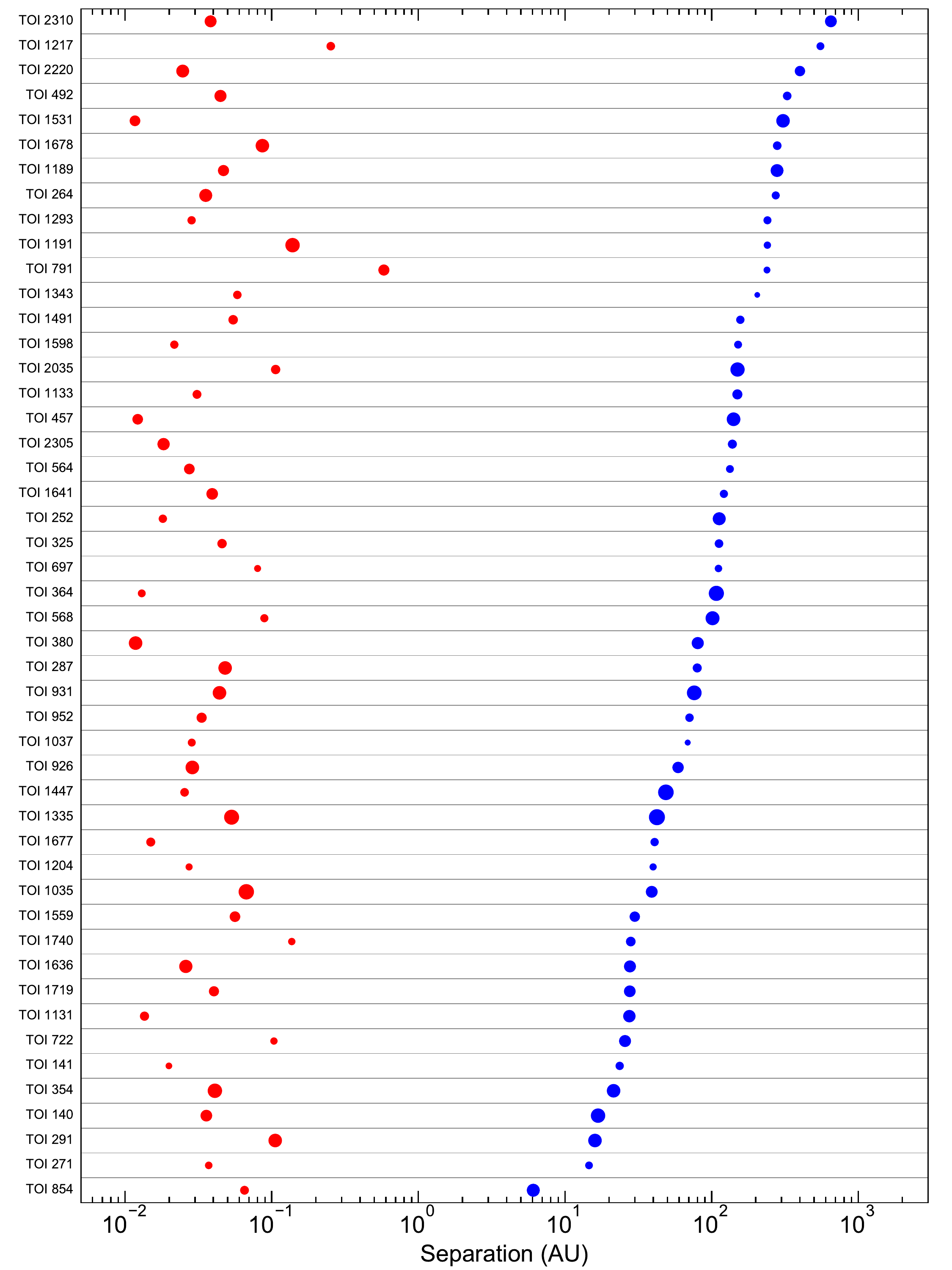}
\caption{Planet and stellar companion separations for the binary exoplanet host systems observed with Gemini. The planets are plotted as red circles (size scaled with the cube root of the planet radius) and the stellar companions are plotted as blue circles (size scaled with the binary mass ratio). The systems are sorted by the stellar companion separation. We see no obvious trends between the two populations, perhaps not unexpected for the very short orbital period planet detections common for TESS.
\label{planet_sepAU}}
\end{figure*}

Figure~\ref{planet_sepAU} shows the separations of the planet candidates and stellar companions for the \numcompexo host binaries in our sample with $\rho<1.2"$ and sorted by companion separation. No corresponding trend with the planet separation is seen, likely because the companions are at least $100\times$ farther from the host stars than the planets and well beyond the limit for dynamically stable circumstellar systems \citep[roughly $5-10 \times$,][]{holman99}.  However, TESS cannot detect longer period planets due to its short observation window, so revisits to these systems during the extended mission will be highly valuable to search for longer period planets and investigate whether their separations are influenced by the stellar companions.

\begin{deluxetable*}{cccc|cccc}
\tablewidth{0pt}
\tablecolumns{8}
\tablecaption{Corrected Radii of Planet Candidates in Close Binary Systems
\label{planettable} }
\tablehead{
\colhead{Planet} 
& \colhead{Correction } 
& \colhead{Apparent $R_{pl}$ }
& \colhead{Corrected $R_{pl}$ } 
& \colhead{Planet} 
& \colhead{Correction } 
& \colhead{Apparent $R_{pl}$  }
& \colhead{Corrected $R_{pl}$ } 
\\[-5pt]
\colhead{Candidate} 
& \colhead{ Factor } 
& \colhead{($R_\oplus$)} 
& \colhead{($R_\oplus$)} 
& \colhead{Candidate} 
& \colhead{ Factor } 
& \colhead{($R_\oplus$)} 
& \colhead{ ($R_\oplus$)} 
}
\startdata 
TOI  140.01 & 1.282 & 12.74 & 16.33  &  TOI 1037.01 & 1.005 &  3.29 &  3.31 \\[-2pt]
TOI  141.01 & 1.007 &  1.66 &  1.67  &  TOI 1039.01 & 1.053 &\nodata&\nodata\\[-2pt]
TOI  224.01 & 1.182 & 14.75 & 17.44  &  TOI 1131.01 & 1.103 &  5.56 &  6.13 \\[-2pt]
TOI  235.01 & 1.119 &\nodata&\nodata &  TOI 1133.01 & 1.038 &  4.86 &  5.05 \\[-2pt]
TOI  252.01 & 1.114 &  3.90 &  4.35  &  TOI 1189.01 & 1.109 & 10.76 & 11.93 \\[-2pt]
TOI  264.01 & 1.008 & 18.25 & 18.40  &  TOI 1191.01 & 1.005 & 25.36 & 25.50 \\[-2pt]
TOI  271.01 & 1.005 &  2.66 &  2.68  &  TOI 1192.01 & 1.034 & 11.65 & 12.04 \\[-2pt]
TOI  287.01 & 1.013 & 20.41 & 20.68  &  TOI 1204.01 & 1.005 &  2.12 &  2.13 \\[-2pt]
TOI  291.01 & 1.132 & 20.52 & 23.23  &  TOI 1217.01 & 1.005 &  4.34 &  4.36 \\[-2pt]
TOI  309.01 & 1.178 &\nodata&\nodata &  TOI 1293.01$^*$& 1.000 &  3.86 &  3.86\\[-2pt]
TOI  325.01 & 1.012 &  5.95 &  6.02  &  TOI 1335.01 & 1.411 & 28.83 & 40.67 \\[-2pt]
TOI  354.01 & 1.178 & 25.20 & 29.68  &  TOI 1343.01 & 1.004 &  4.06 &  4.08 \\[-2pt]
TOI  364.01 & 1.312 &  3.13 &  4.11  &  TOI 1447.01 & 1.351 &  4.43 &  5.99 \\[-2pt]
TOI  380.01 & 1.079 & 20.57 & 22.19  &  TOI 1491.01 & 1.008 &  6.24 &  6.29 \\[-2pt]
TOI  457.01$^*$& 1.000 &  9.12 &  9.12&  TOI 1531.01 & 1.100 &  9.44 & 10.38 \\[-2pt]
TOI  487.01 & 1.147 &  2.73 &  3.13  &  TOI 1544.01 & 1.018 &\nodata&\nodata\\[-2pt]
TOI  487.02 & 1.147 &\nodata&\nodata &  TOI 1559.01 & 1.074 &  9.20 &  9.88 \\[-2pt]
TOI  492.01 & 1.014 & 14.29 & 14.49  &  TOI 1565.01 & 1.212 & 23.87 & 28.93 \\[-2pt]
TOI  564.01 & 1.005 &  9.36 &  9.41  &  TOI 1598.01 & 1.006 &  3.78 &  3.80 \\[-2pt]
TOI  568.01 & 1.214 &  3.55 &  4.31  &  TOI 1636.01 & 1.056 & 18.98 & 20.05 \\[-2pt]
TOI  680.01 &\nodata&  5.74 &\nodata &  TOI 1641.01 & 1.037 & 12.54 & 13.00 \\[-2pt]
TOI  697.01 & 1.004 &  2.03 &  2.04  &  TOI 1677.01 & 1.008 &  5.09 &  5.13 \\[-2pt]
TOI  722.01 & 1.122 &  2.38 &  2.67  &  TOI 1678.01 & 1.037 & 20.34 & 21.10 \\[-2pt]
TOI  791.01 & 1.003 & 10.68 & 10.71  &  TOI 1719.01 & 1.077 &  7.83 &  8.43 \\[-2pt]
TOI  798.01 & 1.321 &\nodata&\nodata &  TOI 1740.01 & 1.018 &  2.46 &  2.51 \\[-2pt]
TOI  854.01 & 1.152 &  4.80 &  5.53  &  TOI 2035.01 & 1.298 &  5.90 &  7.66 \\[-2pt]
TOI  926.01 & 1.065 & 21.00 & 22.37  &  TOI 2220.01 & 1.036 & 17.85 & 18.49 \\[-2pt]
TOI  931.01 & 1.277 & 20.69 & 26.42  &  TOI 2305.01 & 1.013 & 15.21 & 15.40 \\[-2pt]
TOI  952.01 & 1.019 &  7.84 &  7.99  &  TOI 2310.01 & 1.091 & 13.68 & 14.93 \\[-2pt]
TOI 1035.01 & 1.094 & 31.50 & 34.45  &              &       &       &       \\[-2pt]
\enddata  
\tablenotetext{*}{The companion for this TOI is listed as a nearby star in ExoFOP, so its flux has already been accounted for when calculating the planet radius.}
\end{deluxetable*} 

\section{Conclusions}\label{sect-con}
We observed \numobstoi TOI stars using speckle interferometry at Gemini in search of stellar companions to validate and characterize the exoplanets and their stellar environments. We detected \numcompexoall companions around exoplanet candidate host TOI's and \numcompfp companions around false positive TOI's and present the binary parameters for each system. The magnitude differences of exoplanet host binaries can be used to correct for transit dilution when calculating the exoplanet radii and density \citep{ciardi15}, but additional observations may be needed to determine around which stellar component the planet is actually orbiting \citep[e.g.,][]{howell19}.

We then combined our sample of exoplanet host binaries with those found with speckle interferometry at WIYN \citep{howell21} in order to test the predictions of several different binary period distributions. Our results are consistent with past studies \citep{kraus16, fontanive19, ziegler20, ziegler21, howell21} in showing that the binary exoplanet host period distribution is narrower and peaks at a larger value than that of field binaries, but our lack of very close companions ($<100$~AU) is less dramatic than found by these previous studies due to our improved angular resolution and sensitivity to faint companions. Furthermore, our sample clearly shows the observational bias against detecting small planets in transit surveys like TESS due to the third light from a stellar companion. Both the host star multiplicity and this observational bias must be taken into account when calculating the  occurence rates of Earth-sized planets \citep{ciardi15, furlan17, saval20}. We did not find any additional correlations in the properties of the planets or companions in our sample.

Finally, understanding how binary companions affect the formation, evolution, and survival of exoplanets is an important component to understanding planet formation overall. This work focused on the impact of the binary separation, but the relative inclination between the planet and companion, the eccentricity of the binary orbit, and the mass ratio are other important factors  \citep[e.g.,][]{holman99, quintana02, jc15}. We are currently monitoring exoplanet host binaries with shorter periods to resolve the visual and spectroscopic orbits and determine each system's orbital parameters \citep[e.g.,][]{colton21}. We plan to test each system's dynamical stability and potential for habitable planets in order to investigate how these other binary properties affect planet formation and survival.

\acknowledgments
\small{
The authors would like to thank the Gemini staff for their help during our observing runs, as well as the anonymous referee for their thoughtful comments. Kathryn Lester's research is supported by an appointment to the NASA Postdoctoral Program at the NASA Ames Research Center administered by Universities Space Research Association under contract with NASA.This work made use of the High-Resolution Imaging instruments `Alopeke and Zorro, which were funded by the NASA Exoplanet Exploration Program and built at the NASA Ames Research Center by Steve B. Howell, Nic Scott, Elliott P. Horch, and Emmett Quigley. `Alopeke and Zorro were mounted on the Gemini North and South telescopes of the international Gemini Observatory, a program of NSF's NOIRLab, which is managed by the Association of Universities for Research in Astronomy (AURA) under a cooperative agreement with the National Science Foundation on behalf of the Gemini partnership: the National Science Foundation (United States), National Research Council (Canada), Agencia Nacional de Investigación y Desarrollo (Chile), Ministerio de Ciencia, Tecnología e Innovación (Argentina), Ministério da Ciência, Tecnologia, Inovações e Comunicações (Brazil), and Korea Astronomy and Space Science Institute (Republic of Korea). The authors wish to recognize and acknowledge the very significant cultural role and reverence that the summit of Maunakea has always had within the indigenous Hawaiian community. We are most fortunate to have the opportunity to conduct observations from this mountain. This work also made use of the Exoplanet Follow-up Observation Program website, which is operated by the California Institute of Technology, under contract with the National Aeronautics and Space Administration under the Exoplanet Exploration Program.
}
\facilities{Gemini North (`Alopeke), Gemini South (Zorro)} 

\software{
\texttt{astropy} \citep{astropy1, astropy2},~ 
\texttt{Matplotlib} \citep{matplotlib},~ 
\texttt{NumPy} \citep{numpy},~ 
\texttt{SciPy} \citep{scipy}  \\}

\clearpage

\appendix

Our Gemini North \& South observations are listed in Tables~\ref{northlog} and \ref{southlog}, respectively. Each observing log contains the 
TOI number, 
TESS magnitude, 
the inverse of the parallax from Gaia EDR3 \citep{gaia, gaiaEDR3} for the distance,
effective temperature from the TESS Input Catalog \citep{tic81}, 
UT date of observation, 
the $5\sigma$ contrast limits at 0.2\arcsec\ and 1.0\arcsec\ in the blue and red filters, 
and notes for each target. 
The Julian date of each observation can be found in the headers of the archival data hosted on the Gemini Observatory Archive.
If a star's distance was not available in Gaia EDR3, then the value from DR2 or ExoFOP was used. Examples of target notes include false positive identifications and other names of the host star with confirmed exoplanets.


\startlongtable
\begin{deluxetable*}{lcccccccccl}
\tablewidth{0pt}
\tabletypesize{\footnotesize}
\tablecaption{Gemini North Observing Log \label{northlog}    }
\tablehead{ 
\colhead{TOI} & \colhead{TIC}& \colhead{TESS} &  \colhead{Distance} &  \colhead{$T_{\rm eff}$} & \colhead{UT Date}  & \multicolumn{2}{c}{ $\Delta$Mag (562 nm) \ \ \ }  & \multicolumn{2}{c}{$\Delta$Mag (832 nm) \ \ \ } & \colhead{Notes}   
\\ 
\colhead{} &\colhead{} & \colhead{Mag.} & \colhead{(pc)} & \colhead{(K)} & \colhead{(MM/DD/YY)} & \colhead{0.2\arcsec} & \colhead{1.0\arcsec} & \colhead{0.2\arcsec} & \colhead{1.0\arcsec}    }
\startdata 
 139 &   62483237 &   9.4 &   42.4 &  4356 &   10/12/19 &   4.00 &   4.45 &   5.02 &   7.35 		&    \\
 143 &   25375553 &   9.8 &  288.4 &  6371 &     6/9/19 &   4.48 &   5.30 &   5.30 &   8.74 		& WASP-111 \\
 162 &   99493790 &  11.7 &  275.1 &  5405 &     6/9/19 &   3.70 &   3.91 &   5.07 &   7.32 		& APC \\
 198 &   12421862 &   9.9 &   23.8 &  3763 &    10/9/19 &   3.77 &   4.31 &   5.46 &   8.19 		&    \\
     & &       &        &       &     8/4/20 &   4.06 &   4.40 &   5.31 &   7.60      & \\
 223 &  326453034 &  10.3 &  369.7 &  5063 &   10/12/19 &   4.14 &   4.38 &   5.40 &   7.54 		&    \\
 225 &   47525799 &  11.0 &  115.8 &  4658 &   10/13/19 &   3.77 &   4.01 &   5.09 &   7.72 		&    \\
 227 &    2760710 &  13.8 &   52.6 &  2808 &   10/13/19 &   4.06 &   4.21 &   4.90 &   6.32 		&    \\
\enddata 
\end{deluxetable*}

\clearpage

\startlongtable
\begin{deluxetable*}{lcccccccccl}
\tablewidth{0pt}
\tabletypesize{\footnotesize}
\tablecaption{Gemini South Observing Log \label{southlog}    }
\tablehead{ 
\colhead{TOI} & \colhead{TIC} & \colhead{TESS} & \colhead{Distance} & \colhead{$T_{\rm eff}$} & \colhead{UT Date} & \multicolumn{2}{c}{ $\Delta$Mag (562 nm) \ \ \ } & \multicolumn{2}{c}{ $\Delta$Mag (832 nm) \ \ \ } & \colhead{Notes}
\\ 
\colhead{} & \colhead{} & \colhead{Mag.} & \colhead{(pc)} & \colhead{(K)} & \colhead{(MM/DD/YY)} & \colhead{0.2\arcsec} & \colhead{1.0\arcsec} & \colhead{0.2\arcsec} & \colhead{1.0\arcsec}  }
\startdata 
 101 &  231663901 &  12.4 &  372.0 &  5600 &    6/20/19 &   4.45 &   4.79 &   4.85 &   5.68 		& WASP-46 \\
 104 &  231670397 &   9.9 &  317.5 &  6036 &    6/20/19 &   4.84 &   5.20 &   5.32 &   7.09 		& WASP-73 \\
 107 &   92352620 &   9.6 &  210.5 &  6153 &    5/21/19 &   5.36 &   5.64 &   5.60 &   6.71 		& WASP-94 A \\
 109 &   29344935 &  13.2 &  515.6 &  5346 &    7/16/19 &   4.23 &   5.16 &   4.99 &   6.66 		& HATS-14 \\
 112 &  388104525 &  11.6 &  302.4 &  5650 &   11/26/20 &   4.28 &   4.67 &   5.44 &   6.82 		& WASP-119 \\
 116 &  238176110 &  11.0 &  151.2 &  4920 &    7/20/19 &   4.11 &   4.43 &   5.24 &   6.57 		& WASP-91 \\
 130 &  263003176 &   7.4 &   56.9 &  6296 &    7/17/19 &   4.77 &   5.92 &   4.65 &   7.13 		&    \\
 134 &  234994474 &   9.2 &   25.2 &  3794 &    7/14/19 &   4.49 &   4.70 &   5.29 &   7.29 		& L 168-9 \\
\enddata  
\end{deluxetable*}



\begin{thebibliography}{}

\bibitem[Astropy Collaboration et al.(2013)]{astropy1} Astropy Collaboration, Robitaille, T.~P., Tollerud, E.~J., et al.\ 2013, \aap, 558, A33. doi:10.1051/0004-6361/201322068

\bibitem[Astropy Collaboration et al.(2018)]{astropy2} Astropy Collaboration, Price-Whelan, A.~M., Sip{\H{o}}cz, B.~M., et al.\ 2018, \aj, 156, 123. doi:10.3847/1538-3881/aabc4f


\bibitem[Borucki et al.(2010)]{borucki10} Borucki, W.~J., Koch, D., Basri, G., et al.\ 2010, Science, 327, 977  

\bibitem[Ciardi et al.(2015)]{ciardi15} Ciardi, D.~R., Beichman, C.~A., Horch, E.~P., et al.\ 2015, \apj, 805, 16

\bibitem[Cieza et al.(2009)]{cieza09} Cieza, L.~A., Padgett, D.~L., Allen, L.~E., et al.\ 2009, \apjl, 696, L84. doi:10.1088/0004-637X/696/1/L84

\bibitem[Colton et al.(2021)]{colton21} Colton, N.~M., Horch, E.~P., Everett, M.~E., et al.\ 2021, \aj, 161, 21. doi:10.3847/1538-3881/abc9af

\bibitem[Dawson \& Johnson(2018)]{dawson18} Dawson, R.~I. \& Johnson, J.~A.\ 2018, \araa, 56, 175. doi:10.1146/annurev-astro-081817-051853

\bibitem[Duquennoy \& Mayor(1991)]{dm91} Duquennoy, A. \& Mayor, M.\ 1991, \aap, 500, 337

\bibitem[Fontanive et al.(2019)]{fontanive19} Fontanive, C., Rice, K., Bonavita, M., et al.\ 2019, \mnras, 485, 4967. doi:10.1093/mnras/stz671

\bibitem[Furlan et al.(2017)]{furlan17} Furlan, E., Ciardi, D.~R., Everett, M.~E., et al.\ 2017, \aj, 153, 71. doi:10.3847/1538-3881/153/2/71

\bibitem[Gaia Collaboration et al.(2016)]{gaia} Gaia Collaboration, Prusti, T., de Bruijne, J.~H.~J., et al.\ 2016, \aap, 595, A1 

\bibitem[Gaia Collaboration et al.(2020)]{gaiaEDR3} Gaia Collaboration, Brown, A.~G.~A., Vallenari, A., et al.\ 2020, arXiv:2012.01533  

\bibitem[Harris et al.(2020)]{numpy} Harris, C.~R., Millman, K.~J., van der Walt, S.~J., et al.\ 2020, \nat, 585, 357. doi:10.1038/s41586-020-2649-2

\bibitem[Hirsch et al.(2021)]{hirsch20} Hirsch, L.~A., Rosenthal, L., Fulton, B.~J., et al.\ 2021, \aj, 161, 134. doi:10.3847/1538-3881/abd639

\bibitem[Holman \& Wiegert(1999)]{holman99} Holman, M.~J. \& Wiegert, P.~A.\ 1999, \aj, 117, 621. doi:10.1086/300695

\bibitem[Horch et al.(2011)]{horch11} Horch, E.~P., Gomez, S.~C., Sherry, W.~H., et al.\ 2011, \aj, 141, 45 

\bibitem[Horch et al.(2014)]{horch14} Horch, E.~P., Howell, S.~B., Everett, M.~E., et al.\ 2014, \apj, 795, 60

\bibitem[Howell et al.(2011)]{howell11} Howell, S.~B., Everett, M.~E., Sherry, W., et al.\ 2011, AJ, 142, 19  

\bibitem[Howell et al.(2014)]{howell14} Howell, S.~B., Sobeck, C., Haas, M., et al.\ 2014, \pasp, 126, 398 

\bibitem[Howell et al.(2019)]{howell19} Howell, S.~B., Scott, N.~J., Matson, R.~A., et al.\ 2019, \aj, 158, 113. doi:10.3847/1538-3881/ab2f7b

\bibitem[Howell et al.(2021)]{howell21} Howell, S.~B., Matson, R.~A., Ciardi, D.~R., et al.\ 2021, \aj, 161, 164. doi:10.3847/1538-3881/abdec6

\bibitem[Hunter(2007)]{matplotlib} Hunter, J.~D.\ 2007, Computing in Science and Engineering, 9, 90. doi:10.1109/MCSE.2007.55

\bibitem[Jang-Condell(2015)]{jc15} Jang-Condell, H.\ 2015, \apj, 799, 147. doi:10.1088/0004-637X/799/2/147

\bibitem[Kraus et al.(2012)]{kraus12} Kraus, A.~L., Ireland, M.~J., Hillenbrand, L.~A., et al.\ 2012, \apj, 745, 19. doi:10.1088/0004-637X/745/1/19

\bibitem[Kraus et al.(2016)]{kraus16} Kraus, A.~L., Ireland, M.~J., Huber, D., et al.\ 2016, AJ, 152, 8

\bibitem[Malkov(2007)]{malkov07} Malkov, O.~Y.\ 2007, \mnras, 382, 1073. doi:10.1111/j.1365-2966.2007.12086.x

\bibitem[Martin et al.(2014)]{martin14} Martin, R.~G., Nixon, C., Lubow, S.~H., et al.\ 2014, \apjl, 792, L33. doi:10.1088/2041-8205/792/2/L33

\bibitem[Matson et al.(2018)]{matson18} Matson, R.~A., Howell, S.~B., Horch, E.~P., et al.\ 2018, AJ, 156, 31

\bibitem[Matson et al.(2019)]{matson19} Matson, R.~A., Howell, S.~B., \& Ciardi, D.~R.\ 2019, AJ, 157, 211 

\bibitem[Moe \& Kratter(2019)]{moe19} Moe, M. \& Kratter, K.~M.\ 2019, arXiv:1912.01699

\bibitem[Mugrauer \& Michel(2020)]{mugrauer20} Mugrauer, M. \& Michel, K.-U.\ 2020, Astronomische Nachrichten, 341, 996. doi:10.1002/asna.202013825

\bibitem[Musielak et al.(2005)]{musielak05} Musielak, Z.~E., Cuntz, M., Marshall, E.~A., et al.\ 2005, \aap, 434, 355. doi:10.1051/0004-6361:20040238

\bibitem[Pecaut et al.(2012)]{pecaut12} Pecaut, M.~J., Mamajek, E.~E., \& Bubar, E.~J.\ 2012, \apj, 746, 154. doi:10.1088/0004-637X/746/2/154

\bibitem[Pecaut \& Mamajek(2013)]{pecaut13} Pecaut, M.~J. \& Mamajek, E.~E.\ 2013, \apjs, 208, 9

\bibitem[Quintana et al.(2002)]{quintana02} Quintana, E.~V., Lissauer, J.~J., Chambers, J.~E., et al.\ 2002, \apj, 576, 982. doi:10.1086/341808

\bibitem[Raghavan et al.(2010)]{raghavan10} Raghavan, D., McAlister, H.~A., Henry, T.~J., et al.\ 2010, ApJS, 190, 1

\bibitem[Ricker et al.(2015)]{ricker15} Ricker, G.~R., Winn, J.~N., Vanderspek, R., et al.\ 2015, JATIS, 1, 014003

\bibitem[Savel et al.(2020)]{saval20} Savel, A.~B., Dressing, C.~D., Hirsch, L.~A., et al.\ 2020, \aj, 160, 287. doi:10.3847/1538-3881/abc47d

\bibitem[Scott et al.(2018)]{scott18} Scott, N.~J., Howell, S.~B., Horch, E.~P., et al.\ 2018, PASP, 130, 054502

\bibitem[Scott et al.(2021)]{scott21} Scott, N.~J., et al.\ 2021, \textit{in preparation}

\bibitem[Stassun et al.(2018)]{stassun18} Stassun, K.~G., Oelkers, R.~J., Pepper, J., et al.\ 2018, \aj, 156, 102. doi:10.3847/1538-3881/aad050

\bibitem[Stassun et al.(2019)]{tic81} Stassun, K.~G., Oelkers, R.~J., Paegert, M., et al.\ 2019, \aj, 158, 138. doi:10.3847/1538-3881/ab3467

\bibitem[Virtanen et al.(2020)]{scipy} Virtanen, P., Gommers, R., Oliphant, T.~E., et al.\ 2020, Nature Methods, 17, 261. doi:10.1038/s41592-019-0686-2  

\bibitem[Wang et al.(2014)]{wang14} Wang, J., Fischer, D.~A., Xie, J.-W., et al.\ 2014, \apj, 791, 111. doi:10.1088/0004-637X/791/2/111

\bibitem[Winters et al.(2019)]{winters19} Winters, J.~G., Henry, T.~J., Jao, W.-C., et al.\ 2019, \aj, 157, 216

\bibitem[Ziegler et al.(2020)]{ziegler20} Ziegler, C., Tokovinin, A., Brice{\~n}o, C., et al.\ 2020, \aj, 159, 19

\bibitem[Ziegler et al.(2021)]{ziegler21} Ziegler, C., Tokovinin, A., Latiolais, M., et al.\ 2021, arXiv:2103.12076

\end{thebibliography}
\end{document}